\documentclass[12pt, a4paper]{article}
\usepackage{amsmath}
\usepackage{amsthm,keyval}
\usepackage[font=small]{caption}
\usepackage{algorithm, amsfonts, colortbl}
\usepackage{lineno}
\usepackage[noend]{algpseudocode}
\usepackage[top=1.2in, bottom=1.2in, left=1.1in, right=1.1in]{geometry}

\usepackage[round]{natbib}
\usepackage{color, soul, dsfont, tikz}
\usetikzlibrary[decorations.shapes]
\usetikzlibrary[petri]
\sethlcolor{purple}
\usepackage[figuresright]{rotating}
\usepackage{mathrsfs}
\DeclareCaptionStyle{italic}[justification=centering]
 {labelfont={bf},textfont={it},labelsep=colon}
\usepackage{graphicx, psfrag, epsf, textcomp}
\usepackage{enumerate}
\usepackage{natbib}
\usepackage{url, xcolor} 
\usepackage{booktabs, subfig, bm, paralist, mathpazo, tikz, longtable, microtype}

\usepackage{setspace}
\usepackage{mathtools}

\usepackage[pdftex, colorlinks=true]{hyperref}
\definecolor{darkblue}{rgb}{0,0,.6}
\hypersetup{citecolor=darkblue, linkcolor=darkblue, urlcolor=darkblue}
\usepackage{orcidlink}

\newcommand{\blind}{0}

\newcommand{\X}{\mathcal{X}}
\newcommand{\Y}{\mathcal{Y}}

\graphicspath{{plots/}}

\newsavebox\CBox

\newcommand{\pto}{\xrightarrow{p}}

\definecolor{a0}{rgb}{0.0, 0.5, 0.0}
\definecolor{bistre}{rgb}{0.24, 0.17, 0.12}
\definecolor{amethyst}{rgb}{0.6, 0.4, 0.8}
\definecolor{blue-violet}{rgb}{0.54, 0.17, 0.89}
\definecolor{Rcolor}{RGB}{150,160,190}
\definecolor{blush}{rgb}{0.87, 0.36, 0.51}
\definecolor{brightturquoise}{rgb}{0.03, 0.91, 0.87}
\definecolor{burntorange}{rgb}{0.8, 0.33, 0.0}

\newtheorem{definition}{Definition}
\newtheorem{theorem}{Theorem}
\newtheorem{proposition}{Proposition}

\newtheorem{assumption}{Assumption}

\newtheorem{lemma}{Lemma}
\newtheorem{remark}{Remark}

\makeatletter
\let\arxiv@maketitle\maketitle
\let\arxiv@internalmaketitle\@maketitle
\let\arxiv@title\title
\let\arxiv@author\author
\let\arxiv@date\date
\let\arxiv@thanks\thanks
\let\arxiv@and\and
\makeatother

\begin{document}

\def\spacingset#1{\renewcommand{\baselinestretch}%
{#1}\small\normalsize} \spacingset{1}

\if0\blind
{
\title{\bf Local Whittle estimation for distributional time series: \mbox{A regularization framework in Bayes Hilbert space}}
 \author{\normalsize Won-Ki Seo \orcidlink{0000-0001-6629-3027}\thanks{Corresponding author: School of Economics, Level 5, Social Sciences Building, University of Sydney, Sydney, NSW 2050, Australia; Email: won-ki.seo@sydney.edu.au}  \\
    \normalsize School of Economics \\
    \normalsize University of Sydney\\
\\    
    \normalsize Jiazhen Xu \orcidlink{0009-0006-7870-0340} \ and \ Han Lin Shang \orcidlink{0000-0003-1769-6430}\thanks{Shang is grateful for financial support from an Australian Research Council Future Fellowship (FT240100338).}\\
    \normalsize Department of Actuarial Studies and Business Analytics \\
    \normalsize Macquarie University}
\date{}
\maketitle
} \fi

\if1\blind
{
\title{\bf Local Whittle estimation for distributional time series: \mbox{A regularization framework in Bayes Hilbert space}}
\author{}
\maketitle
} \fi

\bigskip

\begin{abstract}
We study local Whittle estimation of long memory in density-valued time series in Bayes Hilbert space. The centered log-ratio (clr) transformation maps densities to a linear Hilbert space, allowing analysis using standard Hilbert space methods. When densities are reconstructed from finite samples, however, small absolute estimation errors can remain large relative to the true density in low-density tails. Taking logarithms can amplify this noise and introduce systematic distortions that affect memory estimation. We develop a framework for regularizing estimated densities before transformation to improve finite-sample performance while preserving the latent memory target. Using the clr representation, we apply local Whittle estimation to principal component scores along an estimated leading direction of the long-run covariance operator. The framework is based on \emph{local Whittle admissibility}, defined by uniform asymptotic equivalence between the feasible and latent-score criteria. We separate density-estimation error from the population perturbation induced by regularization and give sufficient admissibility conditions based on density-estimation rates and the temporal dependence of the error components. Under suitable regularity assumptions, these conditions yield consistency for the latent memory parameter. We study additional tail smoothing, a vanishing density floor, and their combination. Monte Carlo experiments show that these procedures can reduce finite-sample bias relative to unregularized kernel density reconstruction. An application to daily distributions of intraday EUR/USD exchange rate returns illustrates the methodology.\\


\vspace{0.2em}


\end{abstract}

\newpage
\spacingset{1.8}

\section{Introduction}\label{sec:1}

Distributional time series arise in studies of income distributions \citep{KU01}, financial return distributions \citep{KMP+19}, distributions of bid submission times in online auctions \citep{JSP+08}, bicycle availability distributions in Paris \citep{JB26}, and age distributions \citep{SH20,SY26}. We are interested in statistical inference for density-valued time series, with particular emphasis on their temporal persistence. Long memory reflects slowly decaying dependence and has implications for uncertainty quantification and the specification of forecasting models. Studying this persistence requires methods that account for the structure of probability distributions.

The analysis of temporal dependence in distribution-valued time series is closely tied to the geometric representation of the observations. Existing approaches can be broadly classified as intrinsic or extrinsic. Intrinsic approaches operate directly on the underlying metric space of distributions, requiring notions of dependence that do not rely on conventional linear operations \citep{zhou2012measuring, lyons2013distance, jiang2024testing}. The estimation of long memory within an intrinsic framework remains relatively unexplored, with \cite{seo2026quantifying} providing, to the best of our knowledge, the most closely related contribution. In contrast, extrinsic approaches embed probability distributions into a linear function space, allowing functional data analysis techniques to be applied to the transformed observations, with the results mapped back to the original distribution space when necessary \citep{PM16, KMP+19}. Examples include the $\alpha$ transformation \citep{SH25} and the log-quantile-density transformation \citep{PM16}.

We adopt the Bayes Hilbert space framework, which extends the log-ratio geometry of compositional data to probability densities \citep{Aitchison86,egozcue2006hilbert,vandenboogaart2014_bayes}. The associated centered log-ratio (clr) transformation provides an isometric isomorphism to a linear subspace of square-integrable functions, allowing standard functional data analysis methods to be applied while preserving the Bayes Hilbert space geometry. This provides a natural basis for studying temporal dependence through covariance operators, principal component scores, and spectral methods. This Bayes Hilbert space approach has been widely adopted in density data analysis, including principal component analysis (PCA) \citep{hron2016_simplicial}, regression \citep{maier2025_additive}, and time series modeling and forecasting \citep{KMP+19,seo2019_cointegrated}. Local Whittle estimation is an established approach to inference on memory in functional time series \citep{lrs2021_local_whittle,shang2023_sieve,BaekKokoszkaMeng2024,seo_shang2024}. We apply this approach to the clr-transformed density process.

When the densities are unobserved, each must first be reconstructed from a finite sample, introducing estimation noise into the transformed time series. Reconstruction uncertainty is explicitly addressed in the Bayes-space PCA of \citet[Sections~1 and~2]{steyer_greven2026_sparse}, with particular emphasis on sparsely sampled densities. Our concern can arise even when substantial samples yield relatively accurate density estimates. In low-density tails, small absolute errors can remain large relative to the true density and be amplified by the logarithm. Moreover, the nonlinearity of the logarithm can turn mean-zero reconstruction noise into systematic log-density error. Thus, accuracy on the density scale need not ensure accuracy on the clr scale. Related numerical instability and boundary effects near zero are discussed by \citet[Supplementary Material, Section~A.2.2]{lei2023_outlier}.

Such transformed reconstruction errors enter the scores used for local Whittle estimation. \citet[Section~4]{arteche2004_gaussian} examines how the relative magnitude of additive noise and the choice of bandwidth affect the estimator's finite-sample performance. For a signal with a positive memory parameter, ignoring additive short-memory noise uncorrelated with the signal can produce substantial downward finite-sample bias in conventional semiparametric estimators, including local Whittle estimation \citep[Sections~6 and~8]{arteche2006_perturbed}. Intuitively, the noise contributes relatively more at the higher frequencies within the estimation band, making the log-spectrum appear flatter even though the asymptotic memory parameter remains unchanged. This problem has motivated local Whittle methods that explicitly account for noise \citep{hurvich2005_volatility, PNN12}.

In our setting, the errors arise from density reconstruction and can be amplified by the clr transformation in low-density tails. These errors can depend on the latent densities, so their implications for memory estimation require separate analysis. Our simulations in Section~\ref{sec:3} show pronounced downward bias in unregularized memory estimates as persistence increases. These findings motivate modifying the estimated densities before taking logarithms to reduce finite-sample bias while preserving the memory parameter of the unobserved density process. For this purpose, we introduce \emph{local Whittle admissibility}, defined by uniform asymptotic equivalence between the criterion computed from regularized density estimates and its counterpart based on the latent densities.

Our principal contribution is a framework for local-Whittle-admissible density regularization aimed at reducing finite-sample bias while preserving the latent memory target. Regularization may change the estimated clr curves and can itself introduce a population perturbation. We thus separate the remaining density-estimation error from this perturbation and control their contributions relative to the true scores at low frequencies. We establish sufficient conditions for admissibility based on uniform density-estimation rates or the temporal dependence of the error components. Under standard regularity conditions for the latent process, admissibility yields consistency for its memory parameter. This provides a justification for deliberate stabilization of the reconstructed densities while retaining the original asymptotic target.

We complement these general results with three specific regularization procedures: 
\begin{inparaenum}
\item[(i)] additional smoothing in prescribed tail regions, 
\item[(ii)] a vanishing density floor obtained by mixing with the uniform density, and 
\item[(iii)] their combination. 
\end{inparaenum}
Tail smoothing addresses sampling variability where local observations are sparse, while the density floor limits extreme negative log-density values. For each procedure, we give sufficient conditions under which the remaining reconstruction error and any population regularization perturbation are negligible for local Whittle estimation. Monte Carlo experiments further show that these procedures can reduce the finite-sample bias of memory estimates relative to unregularized kernel density reconstruction.

The remainder of the paper is organized as follows. Section~\ref{sec:2} introduces the Bayes Hilbert space representation and the memory parameter. Section~\ref{sec_admissibility} establishes the general conditions for preserving the local Whittle target. Section~\ref{sec_regularization} presents the regularization procedures and their sufficient conditions. Section~\ref{sec:3} reports the simulation study, and Section~\ref{sec:5} presents an empirical illustration of the daily distributions of intraday EUR/USD exchange rate returns. Section~\ref{sec:6} concludes. Proofs and additional simulation results are provided in Sections~\ref{app_proofs} and~\ref{sec:4} of the Supplementary Material, respectively.

\section{Preliminaries}\label{sec:2}

Throughout, we let $\{\X_t:t\in\mathbb Z\}$ be a time series of densities with respect to Lebesgue measure on a common compact support $\mathcal I\subset\mathbb R$ of positive length. Assume that each density is positive almost everywhere.

\subsection{Bayes Hilbert space and the clr transformation}\label{sec:2.1}

Bayes Hilbert spaces extend the Aitchison geometry of compositional data to positive functions and measures, providing a Hilbert space framework for density data \citep{egozcue2006hilbert,vandenboogaart2014_bayes}. We use the ordinary clr transformation associated with the uniform probability measure $P_0$ on $\mathcal I$, where $\mathrm dP_0(u)=|\mathcal I|^{-1}\,\mathrm du$ and $|\mathcal I|$ denotes the length of the support. Let $L^2(P_0)$ denote the space of measurable real-valued functions $g$ on $\mathcal I$ satisfying $\int_{\mathcal I}|g(u)|^2\,\mathrm dP_0(u)<\infty$, with functions equal $P_0$-almost everywhere identified. For a positive function $f$ with $\log f\in L^2(P_0)$, define
\begin{equation*}
\operatorname{clr}f(u)=\log f(u)-\frac{1}{|\mathcal I|}\int_{\mathcal I}\log f(v)\,\mathrm dv.
\end{equation*}
The Bayes Hilbert space $B^2(P_0)$ consists of equivalence classes $[q]$ of positive measurable functions satisfying $\log q\in L^2(P_0)$, where functions differing by a positive multiplicative constant are identified almost everywhere. Vector addition and scalar multiplication correspond to multiplication and real powers of representatives, respectively. Assume that $\log\X_t\in L^2(P_0)$ almost surely for every $t$. The density of the distribution represented by $\X_t$ with respect to $P_0$ is $|\mathcal I|\X_t$, which belongs to the same equivalence class as $\X_t$. Through this identification, we regard $\{\X_t\}$ as a time series taking values in $B^2(P_0)$.

The transformed curves lie in the closed subspace $\mathcal H=L_0^2(P_0)=\{g\in L^2(P_0):\int_{\mathcal I}g\,\mathrm dP_0=0\}$, equipped with the inner product and norm
\begin{equation*}
\langle f,g\rangle=\frac{1}{|\mathcal I|}\int_{\mathcal I}f(u)g(u)\,\mathrm du,\qquad \|f\|^2=\langle f,f\rangle.
\end{equation*}
The Bayes inner product is defined through the inner product of the clr images. With this structure, the clr map is an isometric isomorphism from $B^2(P_0)$ onto $\mathcal H$ \citep{vandenboogaart2014_bayes}. It converts the Bayes vector operations into ordinary addition and scalar multiplication in $\mathcal H$. The uniform density $|\mathcal I|^{-1}$ represents the zero element, since its clr image is zero.

\begin{remark}[Other fixed reference measures]\label{rem_reference}
The framework can also be formulated using another fixed positive reference density $\lambda$, common to all $t$. Write $\mathrm dP_\lambda(u)=\lambda(u)\,\mathrm du$. For a density $f$ with $\log(f/\lambda)\in L^2(P_\lambda)$, define
\begin{equation*}
\operatorname{clr}_\lambda f(u)=\log\left\{\frac{f(u)}{\lambda(u)}\right\}-\int_{\mathcal I}\log\left\{\frac{f(v)}{\lambda(v)}\right\}\,\mathrm dP_\lambda(v).
\end{equation*}
The corresponding analysis uses $L_0^2(P_\lambda)$ with the inner product $\langle f,g\rangle_\lambda=\int_{\mathcal I}fg\,\mathrm dP_\lambda$, and the stated score, direction, and error conditions must hold for this representation. For the primitive conditions in Proposition~\ref{prop:primitive}, also require $\lambda$ to be bounded above and away from zero and replace the density lower bound by the corresponding lower bound on $\X_t/\lambda$. The floor mixtures in Section~\ref{sec_regularization} then use $\lambda$ in place of $|\mathcal I|^{-1}$. Throughout the remainder, we use the ordinary clr transformation and a uniform inner product. The auxiliary reference introduced for $R_1$ determines its smoothing scale and does not change this convention.\qed
\end{remark}

The clr representation has been used to study density-valued time series in several settings (e.g., \citealp{hron2016_simplicial, seo2019_cointegrated, seong2025_functional, maier2025_additive}). We study temporal dependence in the Bayes Hilbert space-valued density process through its clr representation, as described in Section~\ref{sec:2.1}.

\subsection{Score processes and memory parameters}\label{sec:2.2}

Let $\Y_t=\operatorname{clr}\X_t$. For the stationary analysis, assume that $\{\Y_t\}$ is weakly stationary with $\mathbb{E}\|\Y_t\|^2$~$<\infty$. Let $r\in\mathbb N\cup\{\infty\}$, and let $\{v_\ell\}_{\ell=1}^{r}$ be fixed orthonormal directions in $\mathcal H$, initially treated as known. The corresponding population scores are
\begin{equation}
\xi_{t\ell}=\langle\Y_t-\mathbb{E}\Y_t,v_\ell\rangle,\qquad 1\leq\ell\leq r. \label{eqscore}
\end{equation}
We assume that each score process has a spectral density $f_{\xi_\ell}$ satisfying
\begin{equation*}
f_{\xi_\ell}(\omega)\sim G_\ell\omega^{-2d_\ell}\quad\text{as }\omega\downarrow0,\qquad G_\ell>0,\qquad -1/2<d_\ell<1/2.
\end{equation*}
Here, $\sim$ denotes asymptotic equivalence: $a(\omega)\sim b(\omega)$ as $\omega\downarrow0$ means that $a(\omega)/b(\omega)\to1$.
Under this spectral characterization, $d_\ell>0$, $d_\ell=0$, and $d_\ell<0$ correspond to long memory, short memory, and antipersistence, respectively. Nonstationary cases are considered through differencing in Remark~\ref{rem_differencing}, with simulation results reported in Section~\ref{sec:4} of the supplementary material.
We assume that $v_1$ attains the maximal memory order of the clr process among all scalar projections; in particular, $$d_1\geq d_2\geq \ldots \geq d_r.$$
Accordingly, we interpret $d_1$ as the memory parameter of the density-valued process in $B^2(P_0)$. It characterizes the most persistent part of the clr process: for any direction with a smaller memory parameter, the score spectral density is of lower order near zero. Estimating the memory parameter governing such persistent dynamics is an established objective in functional time series analysis \citep{lrs2021_local_whittle, seo_shang2024}. Here, $d_1$ is identified through the low-frequency behavior of $\xi_t$ (i.e., $\xi_{t1}$). Estimation of a direction capturing this maximal memory using a long-run covariance estimator is discussed in Section~\ref{sec_direction}.

\subsection{Local Whittle estimation from latent scores}\label{sec:2.3}

Our target is the memory parameter $d_1\in(-1/2,1/2)$ of the score process $\xi_t$. We first define the local Whittle estimator (see \citealp{robinson1995, lrs2021_local_whittle}) based on the latent scores, which serves as a benchmark for the feasible procedures introduced below.

Let the admissible range $\mathcal D$ for $d$ be a compact interval contained in $(-1/2,1/2)$ with $d_1\in\mathcal D$. Write $\omega_j=2\pi j/n$ for the lowest $m_n$ positive Fourier frequencies, where $m_n\to\infty$ and $m_n/n\to0$. Define the discrete Fourier transform and periodogram of the latent scores by
\begin{equation*}
\mathcal{J}_n^0(\omega_j)=(2\pi n)^{-1/2}\sum_{t=1}^n\xi_t\exp(\mathrm it\omega_j),\qquad I_n^0(\omega_j)=|\mathcal{J}_n^0(\omega_j)|^2,\qquad j=1,\ldots,m_n.
\end{equation*}
The local Whittle objective is
\begin{equation*}
\mathcal L_{m_n}^0(d)=\log S_n^0(d)-\frac{2d}{m_n}\sum_{j=1}^{m_n}\log\omega_j,\qquad S_n^0(d)=\frac{1}{m_n}\sum_{j=1}^{m_n}\omega_j^{2d}I_n^0(\omega_j).
\end{equation*}
The corresponding infeasible estimator is $\widehat d^0=\arg\min_{d\in\mathcal D}\mathcal L_{m_n}^0(d)$.

This estimator is infeasible because the densities $\X_t$ are not directly observed. Each density must be reconstructed from raw observations before the clr curves and scores can be computed. Even when the reconstruction error is small in absolute terms, it can be large relative to $\X_t(u)$ in low-density tails. Taking logarithms can amplify this error and produce systematic distortions, so accuracy on the density scale alone does not ensure accurate memory estimation. The resulting score periodogram contains the true score periodogram, the periodogram of the score error, and cross terms between their Fourier transforms. What matters for local Whittle estimation is the contribution of this error relative to the true score over the frequencies used for estimation. This motivates stabilizing the estimated densities before transformation and assessing the resulting procedure through its local Whittle criterion.

\subsection{Estimated densities and local Whittle estimation}\label{sec:2.4}

We apply local Whittle estimation to a scalar score obtained by projecting each centered clr curve onto the unit-norm function $v_1$ given in \eqref{eqscore}. This function determines how different parts of the curve are weighted in forming the score, and $d_1$ is the memory of the resulting scalar time series. We first treat $v_1$ as known to isolate the effects of density estimation and regularization. When $v_1$ is unknown, a fully feasible implementation also requires estimating this direction. After analyzing these effects, we describe estimation of $v_1$ using a long-run covariance estimator in Section~\ref{sec_direction}, and give conditions under which the resulting memory estimator remains consistent for $d_1$.

Let $\widehat\X_t$ be a preliminary estimator of $\X_t$. We consider deliberately modifying this estimator before applying the clr transformation to stabilize estimated log densities in low-density regions and improve the subsequent memory estimation. We refer to this step as density regularization and use $R$ to index the chosen method. Its validity will be assessed by whether the resulting local Whittle criterion remains asymptotically equivalent to the criterion based on the true scores. Write $\widehat\X_t^R$ for the resulting density estimator and set $\widehat\Y_t^R=\operatorname{clr}\widehat\X_t^R$. Continuing to treat $v_1$ as known, define the estimated score by
\begin{equation*}
\widehat\xi_t^R=\big\langle\widehat\Y_t^R-\frac{1}{n}\sum_{s=1}^n\widehat\Y_s^R,v_1\big\rangle.
\end{equation*}
Its discrete Fourier transform and periodogram at $\omega_j$, $j=1,\ldots,m_n$, are
\begin{equation*}
\mathcal{J}_n^R(\omega_j)=(2\pi n)^{-1/2}\sum_{t=1}^n\widehat\xi_t^R\exp(\mathrm it\omega_j),\qquad I_n^R(\omega_j)=|\mathcal{J}_n^R(\omega_j)|^2.
\end{equation*}
The centering of the sample does not affect $\mathcal{J}_n^R(\omega_j)$ at these nonzero Fourier frequencies. The replacement of the latent periodogram by $I_n^{R}(\omega_j)$ gives the local Whittle objective based on the estimated scores:
\begin{equation}\label{eq_lw_objective}
\mathcal L_{m_n}^R(d)=\log S_n^R(d)-\frac{2d}{m_n}\sum_{j=1}^{m_n}\log\omega_j,\qquad S_n^R(d)=\frac{1}{m_n}\sum_{j=1}^{m_n}\omega_j^{2d}I_n^R(\omega_j).
\end{equation}
The corresponding estimator is 
\[
\widehat d^R=\arg\min_{d\in\mathcal D}\mathcal L_{m_n}^R(d).
\]

The next section introduces \emph{local Whittle admissibility} and states a sufficient condition on the score errors. Together with regularity conditions for the unobserved score process, this condition yields consistency for $d_1$. We then discuss ways to verify the error condition and consider three concrete regularizations: additional tail smoothing with an auxiliary reference ($R_1$, Section~\ref{subsect::R1}); density flooring ($R_2$, Section~\ref{subsect::R2}); and their combination ($R_3$, Section~\ref{subsect::R3}).

\section{Local Whittle admissibility}\label{sec_admissibility}

Density regularization deliberately modifies the inputs to the clr transformation to reduce the influence of unstable tail estimates. To justify these modifications for memory estimation, we control both the remaining reconstruction error and any population perturbation introduced by regularization. Here, $\mathcal L_{m_n}^0$ is the infeasible objective computed from the true scores $\xi_t$ of the unobserved density process, whereas $\mathcal L_{m_n}^R$ is computed from the reconstructed and regularized densities. We state a high-level sufficient condition for the two objectives to be uniformly close and then discuss ways to verify it.

\begin{definition}[Local-Whittle-admissible clr regularization]
A regularization $R$ applied before the clr transformation is local-Whittle-admissible if $\sup_{d\in\mathcal D}\left|\mathcal L_{m_n}^R(d)-\mathcal L_{m_n}^0(d)\right|=o_p(1).$ 
\end{definition}

Under regularity conditions for the true score process, this uniform approximation allows the consistency argument for the infeasible estimator to carry over to $\widehat d^R$. We make these conditions explicit below. To obtain a sufficient condition for admissibility, we separate the contributions of density estimation and population regularization to the score error.

First, let $\overline\Y_n=n^{-1}\sum_{t=1}^n\Y_t$ and define the sample-centered latent score $\xi_{t,n}=\langle\Y_t-\overline\Y_n,v_1\rangle$. Since $\xi_{t,n}=\xi_t-n^{-1}\sum_{s=1}^n\xi_s$ and $\sum_{t=1}^n\exp(\mathrm it\omega_j)=0$ for $j=1,\ldots,n-1$, the population-centered and sample-centered scores have identical discrete Fourier transforms at every nonzero Fourier frequency. Either centering therefore yields the same infeasible local Whittle criterion.

For a regularization $R$, let $\X_t^R$ denote its population density target, and set $\Y_t^R=\operatorname{clr}\X_t^R$. The dependence on $n$ through the regularization parameters is suppressed in the notation. Continuing to treat $v_1$ as known, define the uncentered score errors
\begin{equation*}
\eta_{1,t}^R=\langle\widehat\Y_t^R-\Y_t^R,v_1\rangle,\qquad \eta_{2,t}^R=\langle\Y_t^R-\Y_t,v_1\rangle.
\end{equation*}
Here, $\eta_{1,t}^R$ captures density-estimation error after regularization, clr transformation, and projection onto $v_1$, whereas $\eta_{2,t}^R$ captures the population effect of regularization in the same direction. Define their sample-centered counterparts by $e_{\ell,t}^R=\eta_{\ell,t}^R-n^{-1}\sum_{s=1}^n\eta_{\ell,s}^R$, $\ell=1,2$. The feasible score then admits the exact decomposition
\begin{equation*}
\widehat\xi_t^R=\xi_{t,n}+e_{1,t}^R+e_{2,t}^R.
\end{equation*}
Thus, $e_{1,t}^R$ is the centered density-estimation error and $e_{2,t}^R$ is the centered population regularization remainder. For $\ell=1,2$ and $j=1,\ldots,m_n$, define the Fourier transforms of the score errors and the corresponding weighted periodogram sums by
\begin{equation*}
\mathcal{J}_{\ell,n}^R(\omega_j)=(2\pi n)^{-1/2}\sum_{t=1}^n e_{\ell,t}^R\exp(\mathrm it\omega_j),\qquad S_{\ell,n}^R(d)=\frac{1}{m_n}\sum_{j=1}^{m_n}\omega_j^{2d}|\mathcal{J}_{\ell,n}^R(\omega_j)|^2.
\end{equation*}
Each $S_{\ell,n}^R(d)$ measures the low-frequency contribution of an error component using the same weights $\omega_j^{2d}$ as the periodogram sum $S_n^R(d)$ in \eqref{eq_lw_objective}. Since $e_{\ell,t}^R$ and $\eta_{\ell,t}^R$ differ only by their sample mean, either sequence can be used to compute $\mathcal{J}_{\ell,n}^R(\omega_j)$ at these frequencies. The score decomposition thus gives $\mathcal{J}_n^R(\omega_j)=\mathcal{J}_n^0(\omega_j)+\mathcal{J}_{1,n}^R(\omega_j)+\mathcal{J}_{2,n}^R(\omega_j)$. For the subsequent discussion, define
\begin{equation*}
\mathcal E_{\ell,n}^R:=\sup_{d\in\mathcal D}\frac{S_{\ell,n}^R(d)}{S_n^0(d)},\qquad \ell=1,2.
\end{equation*}
The two quantities, $\mathcal E_{1,n}^R$ and $\mathcal E_{2,n}^R$, measure the relative low-frequency contributions of density estimation and population regularization, respectively.

We first state the benchmark conditions for local Whittle estimation from the true scores. These concern the underlying scalar process $\xi_t$ and its infeasible objective.

\begin{assumption}[Infeasible local Whittle regularity]\label{ass:lw0}
The admissible range $\mathcal D=[\underline d,\overline d]\subset(-1/2,1/2)$ is compact with $d_1\in\mathcal D$ and $\underline d>d_1-1/2$. Suppose that $m_n\to\infty$ and $m_n/n\to0$. Moreover, with $G=G_1>0$,
\begin{equation*}
\sup_{d\in\mathcal D}\left|\frac{S_n^0(d)}{Gm_n^{-1}\sum_{j=1}^{m_n}\omega_j^{2d-2d_1}}-1\right|=o_p(1).
\end{equation*}
\end{assumption}

Assumption~\ref{ass:lw0} collects conventional regularity conditions for scalar local Whittle estimation. On the stated parameter range, its uniform approximation holds for standard stationary linear models, including Gaussian autoregressive fractionally-integrated moving average (ARFIMA) processes, under the usual innovation and spectral regularity conditions; see \citet[Theorem~1 and its proof]{robinson1995} and \citet[Section~2.2]{lrs2021_local_whittle}. The spectral relation $f_{\xi_1}(\omega)\sim G_1\omega^{-2d_1}$ alone does not imply the approximation. The restriction $\underline d>d_1-1/2$ keeps $2(d-d_1)>-1$ throughout $\mathcal D$, allowing the uniform power-sum approximation used in the proof. In particular, this range restriction is automatically satisfied whenever $\overline d-\underline d<1/2$ and $d_1\in\mathcal D$.

The additional condition below concerns the score errors caused by density reconstruction and regularization. We initially impose it at a high level to derive a common admissibility result, before considering sufficient conditions for its verification.

\begin{assumption}[Low-frequency negligibility of the score errors]\label{ass:generic}
For the regularization method $R$ under consideration,
\begin{equation}\label{eq_error_negligibility}
\mathcal E_{1,n}^R+\mathcal E_{2,n}^R=o_p(1).
\end{equation}
\end{assumption}

Since both terms are non-negative, Assumption~\ref{ass:generic} requires $\mathcal E_{1,n}^R=o_p(1)$ and $\mathcal E_{2,n}^R=o_p(1)$. It controls the contribution of each error component relative to the true score over the frequencies used for estimation. It does not require pointwise convergence of the errors or independence between them and the true score.

\begin{theorem}[Admissibility and consistency]\label{thm:consistency}
Under Assumption~\ref{ass:generic},
\begin{equation*}
\sup_{d\in\mathcal D}\left|\mathcal L_{m_n}^R(d)-\mathcal L_{m_n}^0(d)\right|=o_p(1),
\end{equation*}
so $R$ is local-Whittle-admissible. If Assumption~\ref{ass:lw0} also holds, then $\widehat d^R\pto d_1$.
\end{theorem}

For consistency, Assumption~\ref{ass:lw0} ensures that the objective $\mathcal L_{m_n}^0$ based on the true scores assigns uniformly larger values to parameters a fixed distance from $d_1$. Specifically, for every $\epsilon>0$ such that $B_\epsilon=\{d\in\mathcal D:|d-d_1|\geq\epsilon\}$ is nonempty, there is an $\eta_\epsilon>0$ for which
\begin{equation}\label{eq_lw_separation}
\Pr\left[\inf_{d\in B_\epsilon}\{\mathcal L_{m_n}^0(d)-\mathcal L_{m_n}^0(d_1)\}>\eta_\epsilon\right]\to1.
\end{equation}
Indeed, Assumption~\ref{ass:lw0} gives uniform convergence of this contrast to $2(d-d_1)-\log\{1+2(d-d_1)\}$, whose unique minimum is at $d_1$. Thus, \eqref{eq_lw_separation} follows from Assumption~\ref{ass:lw0}. The uniform approximation supplied by Assumption~\ref{ass:generic} preserves this separation for the feasible objective. More generally, the same consistency conclusion holds under Assumption~\ref{ass:generic} whenever \eqref{eq_lw_separation} can be established directly.

We now turn to verification of Assumption~\ref{ass:generic}, continuing to maintain Assumption~\ref{ass:lw0}. The following lemma gives sufficient conditions based on the temporal dependence of an error component. Conditions based on uniform density-estimation rates are developed in Section~\ref{sec_primitive}.

\begin{lemma}[Verification through weaker memory of an error component]\label{lem:lowfreq}
Suppose Assumption~\ref{ass:lw0} holds, and fix $\ell\in\{1,2\}$. Suppose that, for each $n$, the uncentered perturbation $\eta_{\ell,t}^R$ is weakly stationary with $\sup_n\operatorname{Var}(\eta_{\ell,0}^R)<\infty$ and spectral density satisfying
\begin{equation*}
f_{\eta_\ell^R,n}(\omega)=O(|\omega|^{-2d_{\ell,R}}),\qquad -1/2<d_{\ell,R}<d_1,
\end{equation*}
as $\omega\to0$, with the bound holding on a common neighborhood of zero uniformly in $n$. Then
\begin{equation*}
\mathcal E_{\ell,n}^R=O_p\{(m_n/n)^{2(d_1-d_{\ell,R})}\}=o_p(1).
\end{equation*}
\end{lemma}

Taking $\ell=1$ verifies $\mathcal E_{1,n}^R=o_p(1)$ for the density-estimation component, while taking $\ell=2$ verifies $\mathcal E_{2,n}^R=o_p(1)$ for the population regularization component. To establish Assumption~\ref{ass:generic} through this lemma, its conditions must therefore be checked separately for each nonzero error component. An identically zero component requires no further verification. The spectral bound is uniform in $n$ because the regularization parameters may depend on $n$.

The two error components may also be handled using different arguments, applying Lemma~\ref{lem:lowfreq} to one and the uniform error bounds in Proposition~\ref{prop:primitive} to the other. The weaker-memory restriction $d_{\ell,R}<d_1$ is specific to Lemma~\ref{lem:lowfreq}; the uniform error bounds in Proposition~\ref{prop:primitive} provide an alternative way to establish $\mathcal E_{\ell,n}^R=o_p(1)$ without this restriction.


\section{Density regularization methods}\label{sec_regularization}

The regularizations below address the amplification of density-estimation error by the clr transformation. In low-density tails, small absolute estimation errors can remain large relative to the true density and be amplified by the logarithm, so accuracy on the density scale need not translate into accuracy on the clr scale. Moreover, the nonlinearity of the logarithm can turn reconstruction noise into systematic tail distortions, even when the original estimation errors have mean zero. Related numerical instability and boundary effects near zero are discussed by \citet[Supplementary Material, Section~A.2.2]{lei2023_outlier}. Reconstruction uncertainty is also addressed by \citet[Sections~1 and~2]{steyer_greven2026_sparse}, who develop PCA in Bayes spaces with particular emphasis on sparsely sampled densities. We consider a complementary problem in memory estimation that can arise even when densities are reconstructed relatively accurately from substantial samples.

For local Whittle estimation, the concern is how these transformed errors affect the low-frequency behavior of the scores. The finite-sample sensitivity of local Whittle estimation to added noise is examined by \citet[Section~4]{arteche2004_gaussian} in signal-plus-noise models. These considerations motivate deliberately modifying the estimated densities before taking logarithms to reduce the influence of unstable tail estimates. We consider additional tail smoothing, a vanishing density floor, and their combination. Our simulations in Section~\ref{sec:3} show that these modifications can reduce the finite-sample bias of the resulting memory estimates.

The purpose of this stabilization is to improve estimation while preserving the latent memory target. Since modifying the densities can itself perturb the score process, its effect must be controlled together with the remaining reconstruction error. Assumption~\ref{ass:generic} provides a sufficient condition for these effects to be negligible for local Whittle estimation. We examine the three procedures within this framework and give primitive sufficient conditions for their admissibility in Section~\ref{sec_primitive}.

\subsection{\texorpdfstring{$R_{1}$}{R1}: tail smoothing with an auxiliary reference}\label{subsect::R1}

The first regularization is motivated by a change of reference for density smoothing, which suggests wider neighborhoods in the tails. The resulting density estimates are analyzed using the ordinary clr transformation. Let \(A_{t1},\ldots,A_{tN_t}\) denote the raw observations used to estimate \(\X_t\). For bandwidth \(b>0\), write 
\[
\widetilde\X_t(u;b)=N_t^{-1}\sum_{i=1}^{N_t}K_b(u-A_{ti}), 
\]
where \(K_b(x)=b^{-1}K(x/b)\) and $K(\cdot)$ is a non-negative kernel, for the kernel estimate before normalization over \(\mathcal I\); the chosen boundary treatment is understood where needed. The preliminary estimator \(\widehat\X_t\) is obtained by normalizing \(\widetilde\X_t(\cdot;h_n)\), with \(h_n\to0\).

The ordinary clr transformation introduced above uses the uniform reference on \(\mathcal I\). Remark~\ref{rem_reference} explains how densities can also be represented using a different fixed reference. To motivate \(R_1\), suppose we choose a reference that assigns less density to the tails than to the central region. Smoothing over a fixed amount of probability under this reference then corresponds to wider neighborhoods in the tails. We use this reference choice to guide density reconstruction and subsequently apply the ordinary clr transformation to the resulting estimates. Specifically, let \(\mathcal A_n\subset\mathcal I\) be a prescribed union of tail intervals and consider the auxiliary reference density, common to all \(t\),
\begin{equation*}
\rho_n(u)=\frac{\mathbf{1}_{\mathcal I\setminus\mathcal A_n}(u)+\tfrac12\mathbf{1}_{\mathcal A_n}(u)}{Z_n}, \qquad Z_n=|\mathcal I\setminus\mathcal A_n|+\tfrac12|\mathcal A_n|.
\end{equation*}
This reference assigns half as much density to the tails as to the remaining region. Its cumulative coordinate \(F_{\rho_n}(u)=\int_{\inf\mathcal I}^{u}\rho_n(v)\,\mathrm dv\) therefore compresses tail intervals by a factor of two relative to central intervals. Smoothing with bandwidth \(\eta_n=h_n/Z_n\) in this coordinate and transforming the density estimate back gives the local bandwidth \(\eta_n/\rho_n(u)\), equal to \(2h_n\) in the tails and \(h_n\) elsewhere. This bandwidth correspondence applies within regions of constant reference density; kernel contributions spanning their boundaries need not coincide with those of an ordinary kernel density estimation.

Motivated by these two smoothing levels, we implement \(R_1\) directly in the original coordinate and connect them through a convex combination. Let \(s_n:\mathcal I\to[0,1]\) be a deterministic profile common across \(t\), equal to one on \(\mathcal A_n\) and zero in the central region, with a continuous transition between them. Using the same observations for both bandwidths, define
\begin{equation}\label{formula::R1}
Q_{t,n}(u)=[1-s_{n}(u)]\widetilde\X_t(u;h_n)+s_n(u)\widetilde\X_t(u;2h_n), \qquad \widehat\X_t^{R_1}(u)=\frac{Q_{t,n}(u)}{\int_{\mathcal I}Q_{t,n}(v)\,\mathrm dv}.
\end{equation}
Assume that the kernel and boundary treatment are chosen so that \(0<\int_{\mathcal I}Q_{t,n}(v)\,\mathrm dv<\infty\), \(Q_{t,n}\) is positive almost everywhere, and \(\log Q_{t,n}\in L^2(P_0)\), almost surely. The normalized estimator then has a well-defined clr transform.

The baseline component retains the smoothing level associated with the uniform reference, while the wider component implements the tail smoothing suggested by the auxiliary reference. Their convex combination provides a continuous transition between these two levels. Both components use equal observation weights, and normalization is performed after blending. The auxiliary reference changes the smoothing scale used to reconstruct the density. The resulting estimator is then analyzed using the ordinary clr transformation and the uniform inner product.

The motivation is that the logarithm responds to relative density-estimation error. In sparsely sampled tails, \(\X_t(u)\) can be small, so even a small absolute error may produce a large change in the log density, especially when the estimate approaches zero. The centering step in the clr transformation does not remove such local fluctuations. Increasing the tail bandwidth pools observations over a wider neighborhood and can reduce sampling variability, at the cost of additional smoothing bias. The aim is to stabilize the estimated clr curves; Assumption~\ref{ass:generic} requires the resulting score error, including this bias, to be negligible in the local Whittle band.

Both kernel estimates target the same underlying density \(\X_t\), while the wider tail bandwidth changes the bias and variability of the estimator. For the score-error decomposition, we therefore take \(\X_t\) as the population target and use the ordinary clr transformation:
\begin{equation*}
\X_t^{R_1}:=\X_t,\qquad \Y_t^{R_1}=\Y_t,\qquad e_{2,t}^{R_1}=0.
\end{equation*}
The error \(e_{1,t}^{R_1}\) includes both sampling error and finite-bandwidth smoothing bias, including the effect of additional tail smoothing.

\subsection{\texorpdfstring{$R_2$}{R2}: vanishing density floor}\label{subsect::R2}

The second regularization mixes the preliminary estimator with the uniform density on $\mathcal I$ to impose a small positive floor:
\begin{equation*} 
R_2:\quad \widehat\X_t^{R_2}(u)=(1-\epsilon_n)\widehat\X_t(u)+\frac{\epsilon_n}{|\mathcal I|},\qquad 0<\epsilon_n<1,\qquad \epsilon_n\to0.
\end{equation*}
The mixture remains a density and satisfies \(\widehat\X_t^{R_2}(u)\geq\epsilon_n/|\mathcal I|\). Its log density is therefore bounded below by \(\log\{\epsilon_n/|\mathcal I|\}\), limiting large negative values caused by density estimates close to zero. On \(\mathcal I=[0,1]\), the density floor is simply \(\epsilon_n\). Since a fixed floor can alter the population score process, we let \(\epsilon_n\) vanish and control the resulting low-frequency perturbation below.

The population floor target is
\begin{equation*}
\X_t^{R_2}(u)=(1-\epsilon_n)\X_t(u)+\frac{\epsilon_n}{|\mathcal I|}.
\end{equation*}
The resulting clr perturbation determines the floor remainder $e_{2,t}^{R_2}$ in the score-error decomposition. Its low-frequency contribution must be controlled along with the density-estimation error.

\begin{remark}[Positive adjustments]\label{rem_log_offset}
Similar adjustments have been used to stabilize clr transformations of estimated distributions. For example, \citet{lei2023_outlier} preprocess densities $f$ as $f_\alpha=(1-\alpha)f+\alpha$ before applying the clr transformation to mitigate numerical problems caused by near-zero density values. On $[0,1]$, this has the same mixture form as $R_2$. In a related application, \citet{nam_seo2026} impose positive lower bounds on estimated densities. Here, the density floor vanishes with $n$ and is subject to conditions controlling its effect on local Whittle estimation. \qed
\end{remark}

\subsection{\texorpdfstring{$R_3$}{R3}: additional tail smoothing with a vanishing density floor}\label{subsect::R3}

The third regularization applies the same vanishing density floor as \(R_2\) to the tail-smoothed estimator from \(R_1\):
\begin{equation*}
R_3:\quad \widehat\X_t^{R_3}(u)=(1-\epsilon_n)\widehat\X_t^{R_1}(u)+\frac{\epsilon_n}{|\mathcal I|},\qquad 0<\epsilon_n<1,\qquad \epsilon_n\to0.
\end{equation*}
Additional tail smoothing can reduce sampling variability, while the floor ensures \(\widehat\X_t^{R_3}(u)\geq\epsilon_n/|\mathcal I|\).

With \(\X_t^{R_1}=\X_t\) as the population target in the score-error decomposition, \(R_3\) has the same population floor target as \(R_2\):
\begin{equation*}
\X_t^{R_3}=(1-\epsilon_n)\X_t+\frac{\epsilon_n}{|\mathcal I|}=\X_t^{R_2},\qquad e_{2,t}^{R_3}=e_{2,t}^{R_2},\qquad S_{2,n}^{R_3}(d)=S_{2,n}^{R_2}(d).
\end{equation*}
The difference between $R_2$ and $R_3$ lies entirely in their density-estimation errors. Their population floor remainder is identical when they use the same $\epsilon_n$.


\subsection{Primitive sufficient conditions}\label{sec_primitive}

We now give primitive sufficient conditions for Assumption~\ref{ass:generic}. For each method $R$, we require $\mathcal E_{1,n}^R=o_p(1)$. For $R_1$, this is the only requirement because $e_{2,t}^{R_1}=0$. For $R_2$ and $R_3$, we also require $\mathcal E_{F,n}=o_p(1)$, where $\mathcal E_{F,n}:=\mathcal E_{2,n}^{R_2}=\mathcal E_{2,n}^{R_3}$ denotes their common floor contribution. The following proposition verifies these conditions using uniform density-estimation rates and, where applicable, the rate at which the floor vanishes.


\begin{proposition}[Verification through uniform density-estimation rates]\label{prop:primitive}
Suppose Assumption~\ref{ass:lw0} holds, and fix $R\in\{R_1,R_2,R_3\}$. For deterministic $c_n\in(0,1]$, assume
\begin{equation*}
\min_{t\leq n}\operatorname*{ess\,inf}_{u\in\mathcal I}\X_t(u)\geq\frac{c_n}{|\mathcal I|}
\end{equation*}
with probability tending to one. For some deterministic $\delta_n\to0$, we require the uniform error bound for the density estimator used by the selected method:
\begin{equation}\label{eq_kernel_rate}
\begin{cases}
\displaystyle\max_{t\leq n}\|\widehat\X_t-\X_t\|_\infty=O_p(\delta_n), & R=R_2,\\[3pt]
\displaystyle\max_{t\leq n}\|\widehat\X_t^{R_1}-\X_t\|_\infty=O_p(\delta_n), & R\in\{R_1,R_3\}.
\end{cases}
\end{equation}
Define the total clr-error rate by
\begin{equation*}
r_n^R=\begin{cases}
\delta_n/c_n, & R=R_1,\\
(\delta_n+\epsilon_n)/c_n, & R\in\{R_2,R_3\}.
\end{cases}
\end{equation*}
If
\begin{equation}\label{eq_primitive_rate}
n(r_n^R)^2(m_n/n)^{2d_1}\to0,
\end{equation}
then Assumption~\ref{ass:generic} holds. Consequently, $R$ is local-Whittle-admissible and $\widehat d^R\pto d_1$.
\end{proposition}

Here, $\delta_n/c_n$ controls density-estimation error after the logarithm, and the additional term $\epsilon_n/c_n$ controls the population floor perturbation. The fixed support length $|\mathcal I|$ affects only the constants in these bounds. Since $n(m_n/n)^{2d_1}\to\infty$ under Assumption~\ref{ass:lw0}, condition~\eqref{eq_primitive_rate} already implies $r_n^R\to0$. Assumption~\ref{ass:generic} is also a conclusion of the proposition, so it need not be verified separately when the primitive conditions hold. The rate condition is conservative because its proof allows arbitrary temporal concentration of the errors.

\begin{remark}[Separate verification of the floor contribution]\label{rem_separate_floor}
For $R_2$ and $R_3$, the population targets coincide, $\X_t^{R_2}=\X_t^{R_3}=(1-\epsilon_n)\X_t+\epsilon_n|\mathcal I|^{-1}$, so the score errors due to population regularization satisfy $e_{2,t}^{R_2}=e_{2,t}^{R_3}$. Write $\mathcal E_{F,n}:=\mathcal E_{2,n}^{R_2}=\mathcal E_{2,n}^{R_3}$ for their common relative low-frequency contribution. If $\mathcal E_{F,n}=o_p(1)$ is verified separately from the temporal behavior of this error, only the density-estimation error remains to be controlled. Under Assumption~\ref{ass:lw0} and the lower-bound and density-estimation conditions of Proposition~\ref{prop:primitive}, the condition $n(\delta_n/c_n)^2(m_n/n)^{2d_1}\to0$ ensures $\mathcal E_{1,n}^R=o_p(1)$ for the selected method $R\in\{R_2,R_3\}$. Together, these bounds verify Assumption~\ref{ass:generic}, replacing the primitive rate restriction on $\epsilon_n/c_n$ with direct control of the floor error at low frequencies. For example, under Assumption~\ref{ass:lw0}, the bound $\mathbb{E}|\mathcal{J}_{2,n}^{R_2}(\omega_j)|^2\leq C a_n^2\omega_j^{-2d_1}$, uniformly over $j\leq m_n$ for some deterministic $a_n\to0$, gives $\mathcal E_{F,n}=O_p(a_n^2)=o_p(1)$ by the uniform weighted-sum argument in Lemma~\ref{lem:lowfreq}. This bound requires additional control of the error's temporal dependence; a vanishing pointwise floor perturbation alone does not ensure low-frequency negligibility. This alternative establishes admissibility for scores computed along the fixed direction $v_1$. The rate $r_n^R$ used in the subsequent estimated-direction and distributional results must still control the full clr-curve error, including the floor perturbation. \qed
\end{remark}

\begin{remark}[Kernel density estimation]\label{rem_kernel} Proposition~\ref{prop:primitive} uses a generic uniform density-estimation rate $\delta_n$. We now give an explicit rate under standard kernel density assumptions, relating condition~\eqref{eq_primitive_rate} to the within-period sample sizes and the kernel bandwidth. Suppose that, conditional on the latent densities, $A_{t1},\ldots,A_{tN_t}$ are i.i.d. with density $\X_t$ for each $t$.
Let the sample sizes $N_t=N_{t,n}$ be deterministic, set $N_n=\min_{t\leq n}N_t$, and assume $N_t\leq C N_n$ for a fixed $C\geq1$. For each $t$, construct both kernel estimates from the same observations, using common deterministic bandwidths $h_n$ and $2h_n$ across $t$.

Suppose the densities and their derivatives through order two are uniformly bounded by deterministic constants, and $K$ is a non-negative, bounded, Lipschitz kernel of second order; see \citet[Section~3.1]{hansen2008_uniform} for related smoothness and kernel conditions. Use a deterministic boundary correction whose individual kernel contributions are bounded in absolute value by $C/b$, are Lipschitz in $u$ with constant $C/b^2$, and have conditional second moments bounded by $C/b$, uniformly over $t$, $n$, and $b\in\{h_n,2h_n\}$. Assume also that the resulting conditional bias is uniformly $O(h_n^2)$ on $\mathcal I$ at both bandwidths.

If $h_n\to0$ and $\log(nN_n)/(N_nh_n)\to0$, the uniform kernel arguments of \citet[Theorems~2 and~6]{hansen2008_uniform}, applied conditionally and combined with a union bound over $t$ and the two bandwidths, yield
\begin{equation*}
\max_{b\in\{h_n,2h_n\}}\max_{t\leq n}
\big\|\widetilde\X_t(\cdot;b)-\X_t\big\|_\infty
=O_p(\delta_n),
\qquad
\delta_n=
\left\{\frac{\log(nN_n)}{N_nh_n}\right\}^{1/2}+h_n^2.
\end{equation*}
Independence across $t$ is not required for this bound. 
Since $0\leq s_n\leq1$, the mixture satisfies $\max_{t\leq n}\|Q_{t,n}-\X_t\|_\infty=O_p(\delta_n)$ without any smoothness condition on $s_n$. This also gives $\max_{t\leq n}|\int_{\mathcal I}Q_{t,n}(u)\,\mathrm du-1|=O_p(\delta_n)$. Together with the uniform bound on the latent densities, these bounds imply that normalization preserves the rate. The same argument applies to the baseline estimate. Thus, \eqref{eq_kernel_rate} holds for both estimators with the displayed $\delta_n$, which can be substituted into \eqref{eq_primitive_rate} to obtain explicit sufficient conditions for each regularization method. If the boundary treatment has a different uniform bias, replace $h_n^2$ in $\delta_n$ by the corresponding bias bound.\qed
\end{remark}

\subsection{Estimated leading direction}\label{sec_direction}
Let $\widehat v_1^R$ be the unit leading eigenfunction of the long-run covariance estimator $\widehat C_n^R$ computed from the regularized clr curves.
If $\|\kappa_n^{-1}\widehat C_n^R-C\|_{\mathrm{op}}=o_p(1)$ for a positive normalization $\kappa_n$ and a compact non-negative self-adjoint operator $C$ with a simple positive largest eigenvalue and unit eigenfunction $v_1$, then $\|\widehat v_1^R-\operatorname{sgn}(\langle\widehat v_1^R,v_1\rangle)v_1\|=o_p(1)$.
We use this standard eigenfunction consistency result as in \citet{LRS20,lrs2021_local_whittle}.
For the theoretical analysis, define the uncentered score $\widehat\xi_t^{R,\mathrm{est}}=\langle\widehat\Y_t^R,\widehat v_1^R\rangle$, and denote its local Whittle objective and estimator by $\mathcal L_m^{R,\mathrm{est}}$ and $\widehat d^{R,\mathrm{est}}$.
We retain sample centering in applications; for a given estimated direction, it leaves the periodogram unchanged at nonzero Fourier frequencies because $\sum_{t=1}^n\exp(\mathrm it\omega_j)=0$.
Write $\widehat{\mathcal{J}}_n^R(\omega_j)=(2\pi n)^{-1/2}\sum_{t=1}^n\widehat\Y_t^R\exp(\mathrm it\omega_j)$.

\begin{proposition}[Admissibility with an estimated direction]\label{prop_direction}
Suppose $R$ is local-Whittle-admissible for known $v_1$, $\|\widehat v_1^R-\operatorname{sgn}(\langle\widehat v_1^R,v_1\rangle)v_1\|=o_p(1)$, and
\begin{equation}\label{eq_direction_energy}
    E_n^R:=\sup_{d\in\mathcal D}
    \frac{m_n^{-1}\sum_{j=1}^{m_n}\omega_j^{2d}\|\widehat{\mathcal{J}}_n^R(\omega_j)\|^2}{S_n^0(d)}=O_p(1).
\end{equation}
Then $\sup_{d\in\mathcal D}|\mathcal L_m^{R,\mathrm{est}}(d)-\mathcal L_m^0(d)|=o_p(1)$.
Under the separation property \eqref{eq_lw_separation}, $\widehat d^{R,\mathrm{est}}\pto d_1$.
\end{proposition}

Condition~\eqref{eq_direction_energy} requires the total weighted low-frequency variation of the curves to be of at most the same order as that of the target score.
Under \eqref{eq_primitive_rate}, the error bound \eqref{eq_primitive_energy} transfers this condition from the latent curves to the estimated regularized curves.
The proposition concerns objective equivalence and consistency; preservation of a limiting distribution requires stronger rate conditions.

\begin{remark}[Connection with existing results]\label{rem_direction}
Consider the stationary functional linear-process conditions of \citet{lrs2021_local_whittle}, with $-1/2<d_1<1/2$, maintaining the cited moment, spectral, and bandwidth assumptions. The filter coefficients may also be scalar multiples of a fixed bounded operator, as in \citet[Assumption~LW(i) and Appendix~B.2.2]{seo_shang2024}.

Use the Bartlett estimator with weights $1-|h|/b_n$ for $|h|<b_n$, where $b_n\to\infty$ and $b_n=o(n^{1/2})$; if $d_1<0$, also require $b_n^{1-2d_1}r_n^R\to0$. Our estimator is the unnormalized estimator in \citet[equation~(3.1)]{lrs2021_local_whittle} divided by $b_n$, so their normalization $b_n^{2H_0}$, with $H_0=d_1+1/2$, corresponds to $\kappa_n=b_n^{2d_1}$. Their Proposition~2 therefore gives $\|b_n^{-2d_1}\widehat C_n^0-C\|_{\mathrm{op}}=o_p(1)$ for the latent-curve estimator.

Under Proposition~\ref{prop:primitive}, with $r_n^R$ controlling the total centered curve error, \eqref{eq_curve_error} further gives
\begin{equation*}
\left\|b_n^{-2d_1}(\widehat C_n^R-\widehat C_n^0)\right\|_{\mathrm{op}}=O_p\left\{b_n^{1-2d_1}(r_n^R+(r_n^R)^2)\right\}=o_p(1).
\end{equation*}
Thus, if $C$ has a simple positive largest eigenvalue with eigenfunction $v_1$, then $\widehat v_1^R$ consistently estimates $v_1$ up to sign. Since normalization does not change eigenfunctions, no preliminary estimate of $d_1$ is required.

The common scalar filter and \eqref{eq_primitive_energy} also verify \eqref{eq_direction_energy}. Proposition~\ref{prop_direction} therefore yields consistency under the separation property \eqref{eq_lw_separation}, in line with the estimated-direction results of \citet[Theorem~1]{lrs2021_local_whittle} and \citet[Proposition~B.3]{seo_shang2024}. \qed
\end{remark}
The simulation study compares estimates obtained with known and estimated directions to assess the finite-sample effect of estimating $v_1$.

\begin{remark}[Asymptotic normality]\label{rem_lw_clt}
Suppose the conditions of Remark~\ref{rem_direction} hold, $d_1$ lies in the interior of $\mathcal D$, and the asymptotic-normality assumptions of \citet[Theorem~1]{lrs2021_local_whittle} are satisfied. If, in addition, $m_n n(r_n^R)^2({m_n}/{n})^{2d_1}\to0$, where $r_n^R$ controls the total centered curve error in \eqref{eq_curve_error}, including the floor perturbation when present, then
\begin{equation}
\label{eq_lw_clt}\sqrt{m_{n}}\bigl(\widehat d^{R,\mathrm{est}}-d_1\bigr)\xrightarrow{d}N(0,1/4).
\end{equation}
The same limit holds for the known-direction estimator $\widehat d^R$. The additional rate condition makes density estimation and regularization negligible at the $m_n^{-1/2}$ scale; details are given at the end of Section~\ref{app_proofs} of the supplementary material. \qed 
\end{remark}

\begin{remark}[Nonstationary processes and differencing]\label{rem_differencing}
\citet{LRS23} study nonstationary fractionally integrated functional time series with $d_1>1/2$, and \citet{NSS23, NSS26} develop inference methods for functional time series whose nonstationary and stationary components have memory parameters $d_\ell=1$ and $d_\ell=0$, respectively. In our setting, for $1/2<d_1<3/2$, we estimate and regularize the densities as above, then apply local Whittle estimation to the differenced clr curves $\Delta\widehat\Y_t^R=\widehat\Y_t^R-\widehat\Y_{t-1}^R$, $t=2,\ldots,n$. Provided the differenced latent curves satisfy the preceding stationary conditions with target parameter $d_1-1\in(-1/2,1/2)$, we add one to the resulting memory estimate. If the direction is unknown, we estimate it from the differenced clr curves. Under \eqref{eq_curve_error}, the identity $\Delta(\widehat\Y_t^R-\Y_t)=D_{t,n}^R-D_{t-1,n}^R$ shows that differencing preserves the uniform curve-error rate $O_p(r_n^R)$. The preceding consistency and asymptotic-normality conditions therefore apply to the differenced sample with $d_1$ replaced by $d_1-1$. Higher fixed integer differences can be used whenever they bring the target parameter strictly inside $(-1/2,1/2)$ and the resulting latent process satisfies the corresponding stationary conditions.  \qed
\end{remark}

\section{Monte-Carlo simulation studies}\label{sec:3}

We conduct Monte Carlo experiments to evaluate the finite-sample performance of the proposed methodologies. The simulation is designed to reproduce the two sources of error studied in Section~\ref{sec:2}, that is, the latent long-memory variation of the density process and the additional noise introduced by estimating each density from a finite within-curve sample. We consider both the setting in which the leading direction $v_1$ is known and the setting in which $v_1$ is estimated from the observed clr curves.

\subsection{Simulation design}\label{sec:3.1}

Let $\mathcal I=[0,1]$ and use the ordinary clr transformation and uniform inner product defined in Section~\ref{sec:2}. The density-valued time series is generated through a finite-dimensional representation in the clr space. Specifically,
\begin{equation}\label{eq:mc_clr_dgp}
    \Y_t(u)=\mu(u)+\sum_{j=1}^{J}\xi_{tj}v_j(u),
    \qquad t=1,\ldots,n,
\end{equation}
where $J=5$ and $v_1,\ldots,v_J$ are orthonormal in $\mathcal H$, constructed from sine and cosine functions on a grid of 101 equally spaced points and orthonormalized with respect to $\langle\cdot,\cdot\rangle$. The baseline density is $f_0(u)\propto\exp\{-16(u-1/2)^2\}$ and $\mu=\operatorname{clr} f_0$. 

The score processes are mutually independent ARFIMA$(1,d_j,1)$ processes,
\begin{equation*}
(1-\phi_jL)(1-L)^{d_j}\xi_{tj} =(1+\theta_jL)\varepsilon_{tj}, \qquad
\varepsilon_{tj}\stackrel{\text{iid}}{\sim}N(0,\sigma_j^2),
\end{equation*}
with $(d_1,d_2,d_3,d_4,d_5)=\left(d_1,d_1/2,d_1/4,0,0\right)$, $(\phi_1,\ldots,\phi_5)=(0.25,0.15,0.10,0,0)$, $(\theta_1,\ldots,\theta_5)=(-0.20,0.10,0,0,0)$, and $(\sigma_1,\ldots,\sigma_5)=(0.34,0.24,0.17,0.11,0.08)$. A burn-in period of 1200 observations is discarded. The latent density is then obtained through the inverse clr transformation,
\begin{equation}\label{eq:mc_inverse_clr}
\X_t(u)=\frac{\exp\{\Y_t(u)\}}{\int_{\mathcal I}\exp\{\Y_t(v)\}\,\mathrm dv}.
\end{equation}

Here, we consider the setting where the latent densities are not observed directly. Thus, conditional on $\X_t$, we generate
\begin{equation*}
x_{t1},\ldots,x_{tN}\mid \X_t\stackrel{\text{iid}}{\sim}\X_t, \qquad t=1,\ldots,n,
\end{equation*}
and estimate $\X_t$ by a Gaussian kernel density estimator using a Silverman-type bandwidth. Each estimated density is numerically normalized before the clr transformation. We consider $N\in\{100,200,400\}$ in order to examine the effect of within-curve sampling variability.

We compare four feasible procedures. $R_0$ denotes the unregularized kernel density estimator, serving as a baseline. Three proposed density regularization methods in Section~\ref{sec_regularization} are considered. For $R_1$, we apply the bandwidth-blending construction in Section~\ref{subsect::R1}. Based on $R_0$, we blend estimates at the original and doubled bandwidths before normalizing the resulting curve. The weight $s_n(u)$ in \eqref{formula::R1} on the doubled-bandwidth estimate equals one on $[0,0.20]\cup[0.80,1]$, zero on $[0.25,0.75]$, and varies linearly over the two transition regions. For $R_2$, we use the mixture with the uniform density from Section~\ref{subsect::R2}, 
\begin{equation*}
\widehat\X_t^{R_2}(u) = (1-\epsilon_n)\widehat\X_t^{R_0}(u)+\epsilon_n, \qquad \epsilon_n=0.10n^{-0.30}.
\end{equation*}
Finally, $R_3$ applies the same density floor to the $R_1$ weighted kernel estimator as 
\[
\widehat\X_t^{R_3}(u)=(1-\epsilon_n)\widehat\X_t^{R_1}(u)+\epsilon_n.
\]

The simulation design covers the sample size $n$ varying in $\{200,500,1000\}$, the within-curve sample size $N$ varying in $\{100,200,400\}$, and the memory parameter $d_1$ varying in $\{0,0.05,\cdots,0.45\}$ where $d_1=0$ stands for short-memory and higher values of $d_1$ represent increasingly persistent processes. For local Whittle estimation, we take $m_n=\lfloor n^{0.65}\rfloor$. In the main simulations, we set the search interval to $\mathcal D=[d_1-0.49,0.49]$ to satisfy the range restriction in Assumption~\ref{ass:lw0}. Although this choice uses the known simulation value of $d_1$, in applications a preliminary estimate, such as the unregularized local Whittle estimate, can guide the choice of an appropriate interval. In particular, any interval of width less than $1/2$ satisfies the lower-endpoint restriction whenever it contains $d_1$. To assess sensitivity to the use of the true parameter, Section~\ref{sipp sect::unimodal additional simulation} of the supplementary material reports additional experiments using the fixed interval $\mathcal D=[-0.49,0.49]$, which does not depend on $d_1$ and need not satisfy this sufficient restriction. The results are nearly identical to those of the main simulations, indicating that the reported improvements are robust to this broader search interval.

For the long-run covariance estimator constructed from the centered reconstructed clr curves, we follow the Bartlett construction in \citet[equation~(3.1)]{lrs2021_local_whittle} and choose $b_n=\lfloor n^{0.33}\rfloor$. This choice satisfies $b_n\to\infty$ and $b_n=o(n^{1/2})$, as required for their consistency result in Proposition~2. Lag $h$ receives weight $1-h/b_n$ for $h=1,\ldots,b_n-1$, and larger lags receive zero weight. Each design cell is based on 200 Monte Carlo replications.

We report two versions of the estimator. In the known-$v_1$ experiment, feasible clr curves are projected onto the true first basis function $v_1$, and the corresponding infeasible local Whittle estimator based directly on the latent score $\xi_{t1}$ is reported as the Oracle benchmark. In the estimated-$v_1$ experiment, $v_1$ is estimated separately for each regularized clr sample using the leading eigenfunction of its sample covariance operator as described in Section~\ref{sec:2.1}. 

Additional simulation results for different settings, including different baseline densities $f_0(u)$ and nonstationary score processes, are provided in Section~\ref{sec:4} of the supplementary material.

\subsection{Monte Carlo simulation results}\label{sec:3.2}

We first examine the aggregate performance over all combinations of $n\in\{200,500,1000\}$ and $d_1\in\{0,0.05,\cdots,0.45\}$. Table~\ref{table::setting 1} reports average absolute bias, root mean squared error (RMSE), and MAE, using either the true leading direction $v_1$ or its estimate obtained by PCA from the reconstructed clr curves.

\begingroup
\renewcommand{\arraystretch}{1}
\fontsize{7.3pt}{9.4pt}
\selectfont
\setstretch{1.0}
\setlength{\tabcolsep}{0.28in}
\begin{longtable}{@{}clrrrrrr@{}}
\caption{Aggregate Monte Carlo performance. Each entry averages the cell-level absolute bias, RMSE, or MAE equally over all reported $n$ and $d_1$ cells; each cell contains 200 replications. Panel A uses the known $v_1$, whereas Panel B estimates $v_1$. Bias denotes average absolute bias.}\label{table::setting 1} \\
\toprule
& & \multicolumn{3}{c}{Panel A: Known $v_1$} & \multicolumn{3}{c}{Panel B: Estimated $v_1$} \\
\cmidrule(lr){3-5}\cmidrule(lr){6-8}
$N$ & Method & Bias & RMSE & MAE & Bias & RMSE & MAE \\
\midrule
\endfirsthead
\caption[]{Aggregate Monte Carlo performance (continued).} \\
\toprule
& & \multicolumn{3}{c}{Panel A: Known $v_1$} & \multicolumn{3}{c}{Panel B: Estimated $v_1$} \\
\cmidrule(lr){3-5}\cmidrule(lr){6-8}
$N$ & Method & Bias & RMSE & MAE & Bias & RMSE & MAE \\
\midrule
\endhead
\midrule
\multicolumn{8}{r}{\textit{Continued on next page}} \\
\endfoot
\bottomrule
\endlastfoot
100     & $R_0$ & .074 & .115 & .095 & .078 & .118 & .098 \\
        & $R_1$ & .047 & .096 & .077 & .044 & .094 & .075 \\
        & $R_2$ & .030 & .088 & .069 & .024 & .082 & .065 \\
        & $R_3$ & .033 & .089 & .070 & .027 & .083 & .066 \\
\addlinespace
200     & $R_0$ & .061 & .107 & .087 & .070 & .112 & .092 \\
        & $R_1$ & .033 & .090 & .071 & .030 & .087 & .069 \\
        & $R_2$ & .020 & .084 & .066 & .017 & .080 & .063 \\
        & $R_3$ & .024 & .085 & .067 & .020 & .080 & .064 \\
\addlinespace
400     & $R_0$ & .047 & .098 & .078 & .059 & .106 & .086 \\
        & $R_1$ & .022 & .085 & .067 & .018 & .083 & .065 \\
        & $R_2$ & .012 & .082 & .065 & .014 & .079 & .063 \\
        & $R_3$ & .016 & .083 & .065 & .016 & .079 & .063 \\
\end{longtable}
\endgroup

The aggregate results show two main patterns. First, increasing the within-curve sample size $N$ generally reduces all three error measures, but the unregularized estimator $R_0$ retains non-negligible bias even at $N=400$. Second, all three regularizations improve performance relative to $R_0$, with both $R_2$ and $R_3$ performing better than $R_1$. For example, with known $v_1$ and $N=100$, $R_2$ reduces average absolute bias from $0.074$ to $0.030$ and RMSE from $0.115$ to $0.088$.

The unregularized estimator performs worse with an estimated direction than with the known direction, whereas the regularized estimators show relatively small differences between the two settings. For $N=100$, the RMSE of $R_0$ increases from $0.115$ with known $v_1$ to $0.118$ with estimated $v_1$, whereas the corresponding RMSEs for $R_1$, $R_2$, and $R_3$ are slightly smaller with estimated $v_1$.

\begin{figure}[!htb]
\centering
\includegraphics[width=0.9\linewidth]{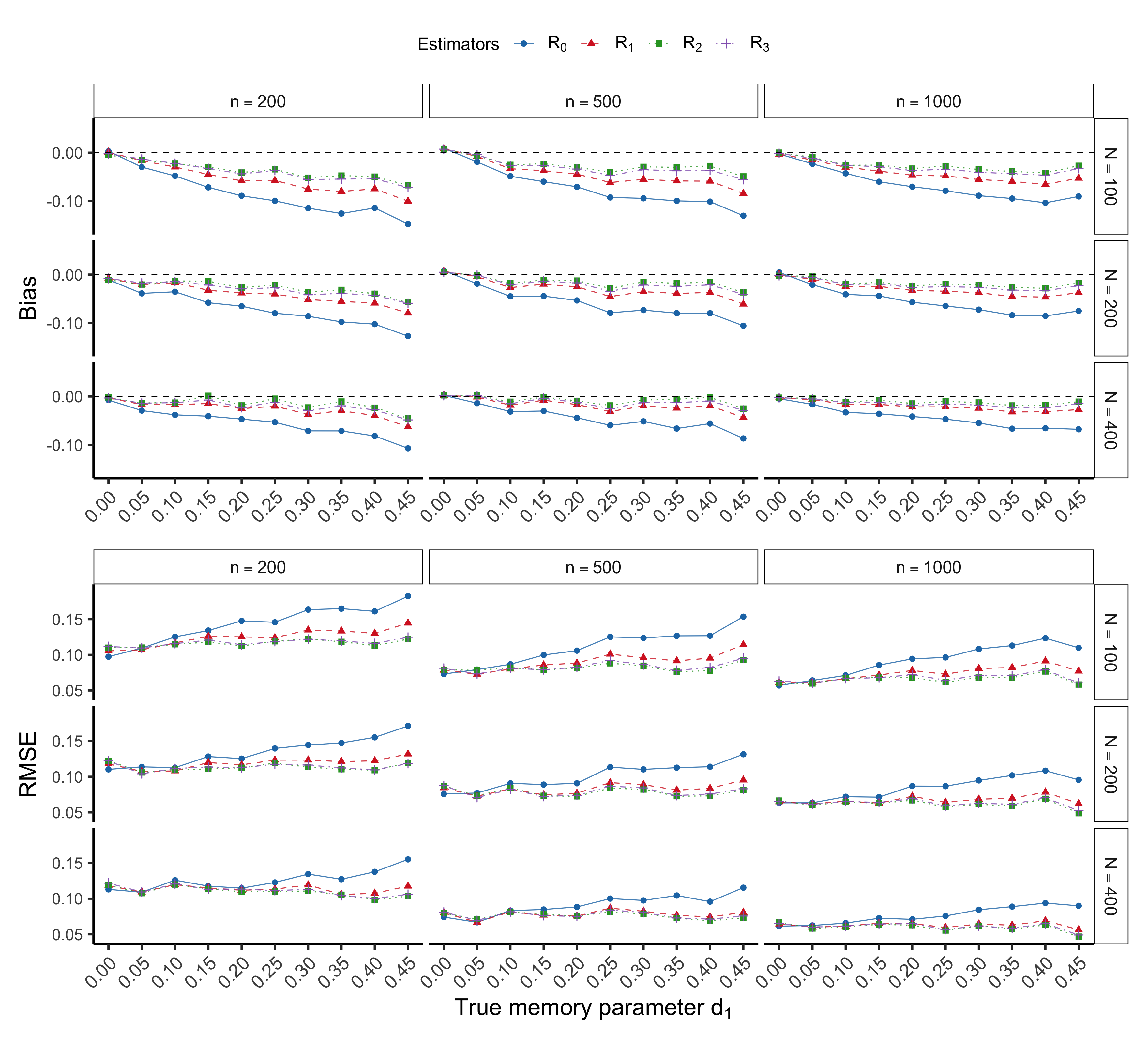}
\caption{\small Bias and RMSE of estimators $R_0$, $R_1$, $R_2$ and $R_3$, for the known $v_1$ setting with sample size $n\in\{200,500,1000\}$, the within-curve sample size $N\in \{100,200,400\}$, and the true long memory parameter $d_1\in\{0,0.05,\cdots,0.45\}$.}\label{fig::sim est known}
\end{figure}

We next examine how performance varies with $n$, $N$, and $d_1$. We begin with the known-$v_1$ experiment, which removes the additional uncertainty associated with estimating the direction. Figure~\ref{fig::sim est known} reports the bias and RMSE of the four estimators and shows a pronounced downward bias for $R_0$ as $d_1$ increases.

Regularizations generally reduce the magnitude of the bias, with $R_2$ and $R_3$ producing similar results. Increasing $N$ also generally reduces the magnitude of the bias and RMSE. By contrast, increasing $n$ alone does not eliminate the distortion when $N$ is held fixed. This persistence is consistent with the simulation design: a longer time series provides more observations for the local Whittle step but does not remove the within-curve density-estimation error.

\begin{figure}[!htb]
\centering
\includegraphics[width=0.9\linewidth]{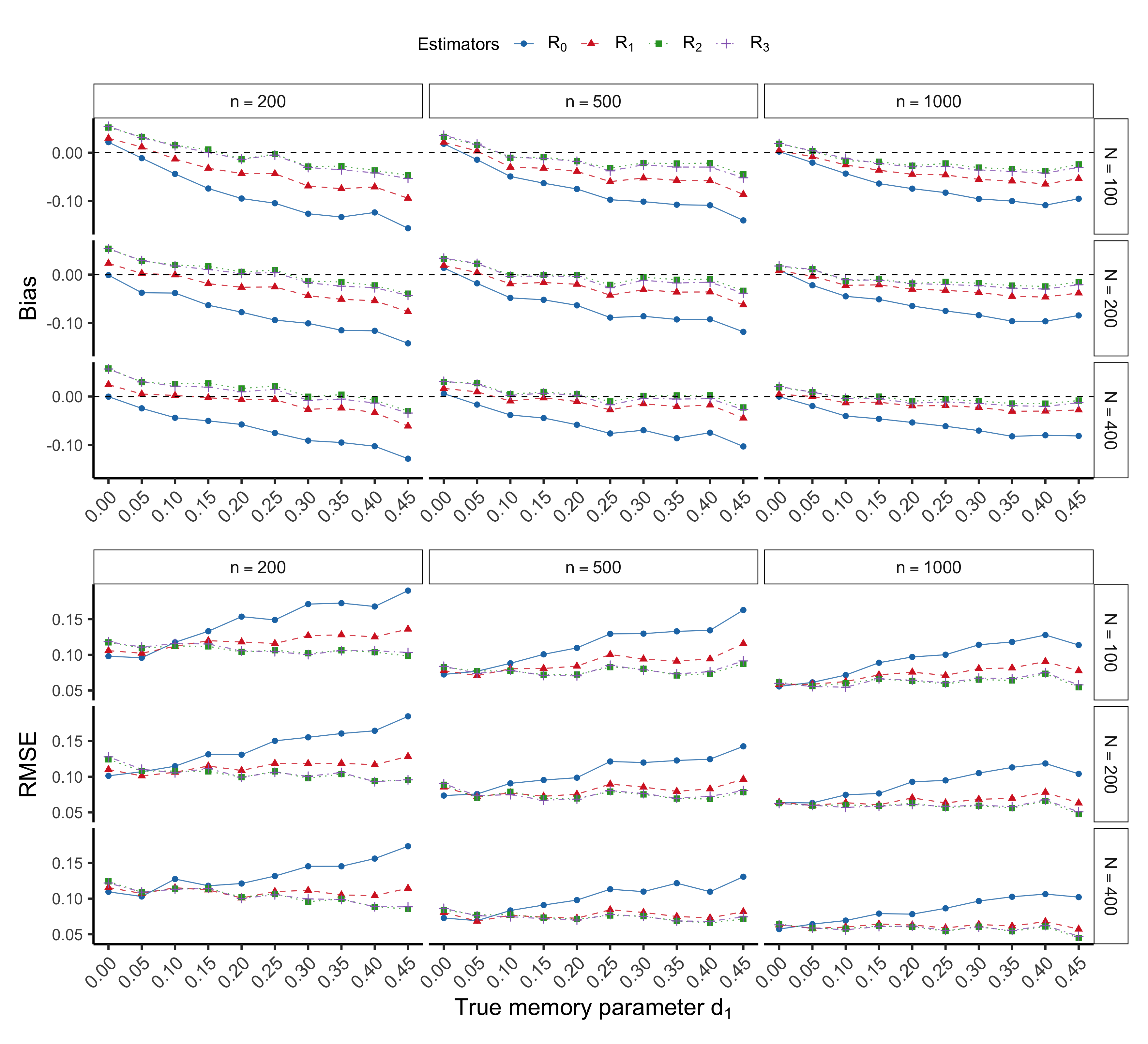}
\caption{\small Bias and RMSE of estimators $R_0$, $R_1$, $R_2$ and $R_3$, for the estimated $v_1$ setting with sample size $n\in\{200,500,1000\}$, the within-curve sample size $N\in \{100,200,400\}$, and the true long memory parameter $d_1\in\{0,0.05,\cdots,0.45\}$.}\label{fig::sim est est}
\end{figure}

We next consider the practically relevant case in which $v_1$ is estimated from the reconstructed clr curves. Figure~\ref{fig::sim est est} shows broadly similar patterns to those in the known-direction experiment. The regularizations retain their improvements when the direction is estimated, consistent with the aggregate findings in Table~\ref{table::setting 1}.

Overall, the results indicate that density-estimation error can limit finite-sample accuracy when the within-curve sample size $N$ is small. The proposed regularizations improve memory estimation with both known and estimated leading directions.

We next assess the finite-sample coverage of confidence intervals for the memory parameter. Motivated by the Hessian-based approximation in \citet[Remark~2]{arteche2025local}, we approximate the variance of $\widehat d$ by $(4S_{m_n})^{-1}$, where $S_{m_n}=\sum_{j=1}^{m_n}(\log j-m_n^{-1}\sum_{k=1}^{m_n}\log k)^2$ and $m_n=\lfloor n^{0.65}\rfloor$. Since $S_{m_n}/m_n\to1$, this approximation agrees asymptotically with the variance implied by Remark~\ref{rem_lw_clt}. At the nominal $95\%$ coverage probability, a confidence interval is 
\[
\left(\widehat d-z_{0.975}(4S_{m_n})^{-1/2},\widehat d+z_{0.975}(4S_{m_n})^{-1/2}\right), 
\]
where $z_{0.975}$ is the $0.975$ quantile of the standard normal distribution. For each simulation cell, empirical coverage is the proportion of the 200 replications in which the interval contains $d_1$.

\begin{figure}[!htb]
\centering
\includegraphics[width=0.9\linewidth]{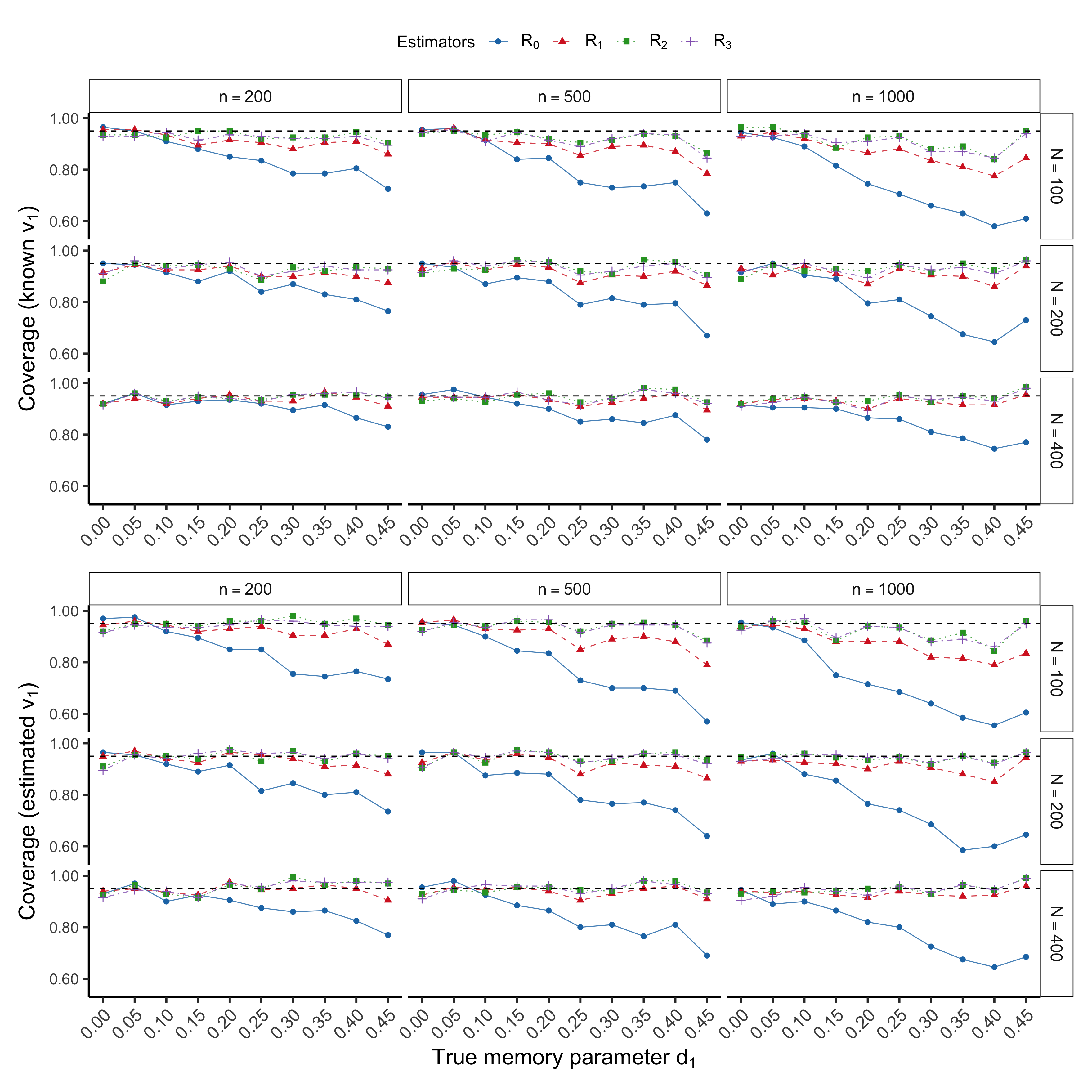}
\caption{\small Empirical coverage probabilities produced by estimators $R_0$, $R_1$, $R_2$ and $R_3$, for either the known or estimated $v_1$ setting with sample size $n\in\{200,500,1000\}$, the within-curve sample size $N\in \{100,200,400\}$, and the true long memory parameter $d_1\in\{0,0.05,\cdots,0.45\}$. The horizontal dashed line stands for the nominal coverage probability of $0.95$.}\label{fig::sim coverage unimodel}
\end{figure}

Figure~\ref{fig::sim coverage unimodel} reports empirical coverage for either the known or estimated $v_1$ setting. The unregularized estimator $R_0$ exhibits substantial undercoverage as the memory parameter increases, particularly when $N$ is small. This pattern is consistent with the downward bias observed in Figure~\ref{fig::sim est known}. All three regularizations generally improve coverage in the more persistent cases. The improvement from $R_1$ is smaller, while $R_2$ and $R_3$ produce similar coverage probabilities that are usually closer to the nominal level. These coverage gains align with the reductions in bias observed in the corresponding estimation results.

\section{Daily distributions of foreign exchange intraday return}\label{sec:5}

Our empirical analysis draws on intraday EUR/USD exchange rate data for 2018, obtained from \url{https://www.kaggle.com/datasets/arashnic/stock-data-intraday-minute-bar/data}. The data consist of closing prices sampled at 5-minute intervals, from which intraday log returns are calculated. To exclude trading days with substantial missing observations, we first determine the most frequently occurring daily observation count and retain days with at least half of this count. We then aggregate the intraday returns for each eligible day into an empirical probability distribution, thereby transforming the original price series into a sequence of distribution-valued observations. The resulting time series comprises $n=259$ daily distributions, each characterizing the intraday return behavior of the EUR/USD exchange rate. We use the ordinary clr transformation with a uniform reference for all four procedures.

Table~\ref{table::real data} reports and long memory estimation produced by $R_0$, $R_1$, $R_2$ and $R_3$, together with the confidence interval. All results indicate that the daily distributions of foreign exchange intraday returns exhibit long memory, consistent with the intrinsic-method findings in Section~5.1 of \cite{seo2026quantifying}.
\begin{table}[!htb]
\centering
\renewcommand{\arraystretch}{1}
\caption{Long memory estimates and 95\% confidence intervals for daily distributions of foreign exchange intraday return.}\label{table::real data}
\fontsize{7.3pt}{9.4pt}\selectfont
\setstretch{1.0}
\tabcolsep 0.5in
\begin{tabular}{@{}lrrr@{}}
\toprule
Method & $\widehat d^{R,\mathrm{est}}$  & Confidence interval \\
\midrule
R0 & .397  & [.209, .585] \\
R1 & .402 & [.214, .591] \\
R2 & .417  & [.229, .605] \\
R3 & .419  & [.231, .607] \\
\bottomrule
\end{tabular}
\end{table}

We then apply the test statistic in Section~3.2 of \cite{SS26} to test whether the clr time series has $d_1=0$. The test statistics based on the $R_0$, $R_1$, $R_2$, and $R_3$ procedures reject the null hypothesis $d_1=0$, supporting the memory estimate in Table~\ref{table::real data}. We then apply the fractional difference based on the memory estimate and apply the test statistic to the fractional-differenced clr time series. The test statistic applied to the fractional-differenced clr time series using all the $R_0$, $R_1$, $R_2$, and $R_3$ procedures does not reject the null hypothesis.

\section{Conclusion}\label{sec:6}


We develop a regularization framework for local Whittle estimation of long memory in density-valued time series observed through finite samples. Within the Bayes Hilbert space representation, the framework addresses the amplification of density-reconstruction errors by the centered log-ratio transformation, particularly in low-density regions. It provides a theoretical justification for stabilizing reconstructed densities while preserving the memory parameter of the latent density process.

The main theoretical contribution is a set of sufficient conditions for local Whittle admissibility, formulated through uniform asymptotic equivalence between the criteria based on regularized density estimates and latent densities. By separating the remaining reconstruction error from the population perturbation induced by regularization, we identify conditions under which both contributions are negligible relative to the latent scores at low frequencies. Together with standard regularity conditions for the latent process, these results establish consistency of the feasible memory estimator. They also clarify that the suitability of regularization depends on its effect on the low-frequency behavior of the transformed process.

We develop three regularization procedures: tail smoothing, mixing with the uniform density via a vanishing weight, and their combination. Monte Carlo experiments show that these procedures can reduce the finite-sample bias associated with unregularized kernel density reconstruction. The empirical illustration applies the proposed framework to daily distributions of intraday EUR/USD exchange rate returns, where each density is reconstructed from within-day observations.

The methodology admits several extensions, of which we mention two.
\begin{inparaenum}[1)]
\item We have focused on estimating the long-memory parameter, a central input to the functional autoregressive fractionally integrated moving average model. A natural next step is to use the estimated memory to specify and fit such a model, and to assess whether accounting for persistence improves density forecasts.
\item A further extension concerns memory reduction through nontrivial linear combinations of density-valued series in Bayes Hilbert space, a phenomenon known as fractional cointegration. Future work could examine how to adapt the present local Whittle approach to estimate the reduced memory order and whether density regularization can reduce the finite-sample bias of the resulting estimates.\\\bigskip
\end{inparaenum}

\noindent\textbf{Supplementary Material:} The supplement contains the proofs of the main results and additional simulation studies. \bigskip  

\noindent\textbf{Data availability statement:} Data and code are available in the \href{https://github.com/d5rhtkp4kc-alt/LW-Dist}{GitHub repository}. \bigskip

\noindent\textbf{Statement on AI Use:} We used AI tools (ChatGPT 5.6 Sol and Claude Opus 5) for language editing; and for programming code, we used them for minor syntax checking, visual alignment, layout adjustments, and debugging. All outputs were thoroughly verified and finalized by the authors. All scientific results are our own.

\appendix

\section*{Supplementary Material}

This supplementary material is organized into two sections. Section~\ref{app_proofs} contains the proofs of main results. Section~\ref{sec:4} provides additional simulation results.

\section{Proofs of main results}\label{app_proofs}

\begin{proof}[\textbf{Proof of Theorem~\ref{thm:consistency}}]
At the nonzero frequencies, the score decomposition gives $\mathcal{J}_n^R=\mathcal{J}_n^0+\mathcal{J}_{1,n}^R+\mathcal{J}_{2,n}^R$. Write $\mathcal{J}_{b,n}=\mathcal{J}_{1,n}^R+\mathcal{J}_{2,n}^R$ and $S_{b,n}(d)=m_n^{-1}\sum_{j=1}^{m_n}\omega_j^{2d}|\mathcal{J}_{b,n}(\omega_j)|^2$. Then $S_{b,n}(d)\leq2S_{1,n}^R(d)+2S_{2,n}^R(d)$. Expanding $|\mathcal{J}_n^0(\omega_j)+\mathcal{J}_{b,n}(\omega_j)|^2$ gives
\begin{equation*}
\begin{aligned}
|S_n^R(d)-S_n^0(d)|&\leq S_{b,n}(d)+2\left|\frac{1}{m_n}\sum_{j=1}^{m_n}\omega_j^{2d}\mathcal{J}_n^0(\omega_j)\overline{\mathcal{J}_{b,n}(\omega_j)}\right|\leq S_{b,n}(d)+2\{S_n^0(d)S_{b,n}(d)\}^{1/2},
\end{aligned}
\end{equation*}
where the second inequality follows from Cauchy--Schwarz applied to the weighted cross term. Dividing by $S_n^0(d)$ yields
\begin{equation*}
\left|\frac{S_n^R(d)}{S_n^0(d)}-1\right|
\leq\frac{S_{b,n}(d)}{S_n^0(d)}
+2\left\{\frac{S_{b,n}(d)}{S_n^0(d)}\right\}^{1/2}.
\end{equation*}
Assumption~\ref{ass:generic} makes the right-hand side $o_p(1)$ uniformly over $\mathcal D$. Since $\mathcal L_{m_n}^R(d)-\mathcal L_{m_n}^0(d)=\log\{S_n^R(d)/S_n^0(d)\}$, continuity of the logarithm around one proves admissibility.

Since $\underline d>d_1-1/2$, the standard power-sum
approximation used in the proof of Theorem 1 of \citet[equations (3.7) and (3.8)]{robinson1995} implies $\sup_{d\in\mathcal D}
|\frac{1}{m_n}\sum_{j=1}^{m_n}\left({j}/{m_n}\right)^{2(d-d_1)}-({1+2(d-d_1)})^{-1}|\to0$ and also $m_n^{-1}\sum_{j=1}^{m_n}\log(j/m_n)\to-1$. These results determine the limit of the objective difference. Indeed, Assumption~\ref{ass:lw0} and $\omega_j=\omega_{m_n}(j/m_n)$ give $\mathcal L_{m_n}^0(d)-\mathcal L_{m_n}^0(d_1)=\log\{m_n^{-1}\sum_{j=1}^{m_n}(j/m_n)^{2(d-d_1)}\}-2(d-d_1)m_n^{-1}\sum_{j=1}^{m_n}\log(j/m_n)+o_p(1)$ uniformly over $d\in\mathcal D$ under the employed assumptions, with the terms involving $\log\omega_{m_n}$ canceling. The power-sum approximation gives the limit $-\log\{1+2(d-d_1)\}$ for the first term, while the log-sum approximation gives $2(d-d_1)$ for the second. Consequently,
\begin{equation*}
\sup_{d\in\mathcal D}\left|\mathcal L_{m_n}^0(d)-\mathcal L_{m_n}^0(d_1)-\left[2(d-d_1)-\log\{1+2(d-d_1)\}\right]\right|=o_p(1).
\end{equation*}
The limiting contrast is continuous on $\mathcal D$ and uniquely minimized at $d_1$, which proves \eqref{eq_lw_separation}. Fix $\epsilon>0$ with $B_\epsilon$ nonempty. Combining this separation with admissibility gives, with probability tending to one,
\begin{equation*}
\inf_{d\in B_\epsilon}\{\mathcal L_{m_n}^R(d)-\mathcal L_{m_n}^R(d_1)\}>\eta_\epsilon/2.
\end{equation*}
The minimizer $\widehat d^R$ thus lies outside $B_\epsilon$ with probability tending to one. This proves consistency, and also shows that directly establishing \eqref{eq_lw_separation} is sufficient for the final step.
\end{proof}

\begin{proof}[\textbf{Proof of Lemma~\ref{lem:lowfreq}}]
At nonzero frequencies, the transforms of $e_{\ell,t}^R$ and $\eta_{\ell,t}^R$ coincide. By \citet[p.~255, equation~(8)]{sykulski2019_debiased}, $\mathbb{E}|\mathcal{J}_{\ell,n}^R(\omega_j)|^2$ can be expressed as a weighted integral of the spectral density $f_{\eta_\ell^R,n}$ over $[-\pi,\pi]$. We split this integral into the fixed neighborhood of zero on which the spectral bound holds and the remaining frequencies. Over the neighborhood of zero, $f_{\eta_\ell^R,n}(u)\leq C|u|^{-2d_{\ell,R}}$ and $-1/2<d_{\ell,R}<1/2$ imply that the weighted integral is at most $C\omega_j^{-2d_{\ell,R}}$. At the remaining frequencies, the weights are bounded by $C/n$, so the employed uniform variance bound makes this part of the integral at most $C/n$. Since $n^{-1}=O(\omega_j^{-2d_{\ell,R}})$ uniformly over $1\leq j\leq m_n$, adding the two bounds gives
\begin{equation}\label{eq_error_periodogram}
\mathbb{E}|\mathcal{J}_{\ell,n}^R(\omega_j)|^2\leq C\omega_j^{-2d_{\ell,R}},\qquad 1\leq j\leq m_n,\end{equation}
for all sufficiently large $n$, with $C$ independent of $n$ and $j$. 

Since $\underline d>d_1-1/2>d_{\ell,R}-1/2$, the power sums $m_n^{-1}\sum_{j=1}^{m_n}\omega_j^{2d-2d_{\ell,R}}$ are bounded above and below by positive constant multiples of $\omega_{m_n}^{2d-2d_{\ell,R}}$, uniformly over $d\in\mathcal D$. Using $(j/m_n)^{2(d-d_{\ell,R})}\leq(j/m_n)^{2(\underline d-d_{\ell,R})}$ and \eqref{eq_error_periodogram}, we obtain
\begin{equation*}\mathbb{E}\left[\sup_{d\in\mathcal D}\frac{S_{\ell,n}^R(d)}{m_n^{-1}\sum_{j=1}^{m_n}\omega_j^{2d-2d_{\ell,R}}}\right]\leq\frac{C}{m_n}\sum_{j=1}^{m_n}(j/m_n)^{2(\underline d-d_{\ell,R})}=O(1).\end{equation*}
Then Markov's inequality, combined with the above expectation bound, thus gives $S_{\ell,n}^R(d)=O_p\{m_n^{-1}\sum_{j=1}^{m_n}\omega_j^{2d-2d_{\ell,R}}\}$ uniformly over $d\in\mathcal D$. By the approximation given in Assumption~\ref{ass:lw0},
\begin{equation*}\sup_{d\in\mathcal D}\frac{S_{\ell,n}^R(d)}{S_n^0(d)}=O_p\left(\sup_{d\in\mathcal D}\frac{\sum_{j=1}^{m_n}\omega_j^{2d-2d_{\ell,R}}}{\sum_{j=1}^{m_n}\omega_j^{2d-2d_1}}\right)=O_p\{\omega_{m_n}^{2(d_1-d_{\ell,R})}\}.\end{equation*}
The last step follows from $\omega_j^{2(d_1-d_{\ell,R})}\leq\omega_{m_n}^{2(d_1-d_{\ell,R})}$ for every $j\leq m_n$. Since $\omega_{m_n}=2\pi m_n/n\to0$ and $d_{\ell,R}<d_1$, the result follows.
\end{proof}

\begin{proof}[\textbf{Proof of Proposition~\ref{prop:primitive}}]
First, $n(m_n/n)^{2d_1}\to\infty$ under Assumption~\ref{ass:lw0}, so \eqref{eq_primitive_rate} implies $r_n^R\to0$. For the selected method, \eqref{eq_kernel_rate} and the fact that flooring multiplies density-estimation errors by $1-\epsilon_n$ give
\begin{equation*}
\max_{t\leq n}\|\widehat\X_t^R-\X_t^R\|_\infty=O_p(\delta_n).
\end{equation*}
On the event that the assumed lower bound holds, the population targets satisfy $\X_t^R\geq c_n/|\mathcal I|$. For $R_1$ this is immediate; for $R_2$ and $R_3$ it follows from $(1-\epsilon_n)c_n+\epsilon_n\geq c_n$. Since $|\mathcal I|$ is fixed and $\delta_n/c_n\to0$,
\begin{equation*}
\max_{t\leq n}\left\|\frac{\widehat\X_t^R-\X_t^R}{\X_t^R}\right\|_\infty=O_p(\delta_n/c_n)=o_p(1).
\end{equation*}
The bound $|\log(1+x)|\leq2|x|$ for $|x|\leq1/2$ and the contraction property of clr centering in $L^2(P_0)$ imply
\begin{equation*}
\max_{t\leq n}\|\widehat\Y_t^R-\Y_t^R\|=O_p(\delta_n/c_n).
\end{equation*}

For $R_1$, the population clr perturbation is $\Y_t^{R_1}-\Y_t=0$, and hence $e_{2,t}^{R_1}=0$. For $R_2$ and $R_3$, put $q_t=|\mathcal I|\X_t$. Their common log-ratio perturbation is $g_{t,n}:=\log(\X_t^{R_2}/\X_t)=\log(\X_t^{R_3}/\X_t)=\log\{1+\epsilon_n(q_t^{-1}-1)\}$. The corresponding population clr perturbation is $\Y_t^{R_2}-\Y_t=\Y_t^{R_3}-\Y_t=g_{t,n}-|\mathcal I|^{-1}\int_{\mathcal I}g_{t,n}(u)\,\mathrm du$. On the same event, $\|\epsilon_n(q_t^{-1}-1)\|_\infty\leq\epsilon_n/c_n=o(1)$, so $\max_{t\leq n}\|g_{t,n}\|_\infty\leq2\epsilon_n/c_n$ for all sufficiently large $n$. Since
\begin{equation*}
\Y_t^R-\Y_t=g_{t,n}-\frac{1}{|\mathcal I|}\int_{\mathcal I}g_{t,n}(u)\,\mathrm du,
\end{equation*}
clr centering gives $\max_{t\leq n}\|\Y_t^R-\Y_t\|=O_p(\epsilon_n/c_n)$. Temporal centering increases either component bound by at most a factor of two.

Consequently, the total centered curve error $D_{t,n}^R=\widehat\Y_t^R-\Y_t-n^{-1}\sum_{s=1}^n(\widehat\Y_s^R-\Y_s)$ satisfies
\begin{equation}\label{eq_curve_error}
\max_{t\leq n}\|D_{t,n}^R\|=O_p(r_n^R).
\end{equation}
For $\mathcal{J}_{D,n}^R(\omega_j)=(2\pi n)^{-1/2}\sum_{t=1}^nD_{t,n}^R\exp(\mathrm it\omega_j)$, the triangle inequality yields
\begin{equation*}
\max_{j\leq m_n}\|\mathcal{J}_{D,n}^R(\omega_j)\|^2\leq\frac{n}{2\pi}\max_{t\leq n}\|D_{t,n}^R\|^2=O_p\{n(r_n^R)^2\}.
\end{equation*}
Assumption~\ref{ass:lw0} and $\underline d>d_1-1/2$ give $S_n^0(d)\geq c\,\omega_{m_n}^{2d-2d_1}$ uniformly over $\mathcal D$ with probability tending to one, for some $c>0$. Since $\underline d>-1/2$, the power sums also give
\begin{equation*}
\sup_{d\in\mathcal D}\frac{m_n^{-1}\sum_{j=1}^{m_n}\omega_j^{2d}}{S_n^0(d)}=O_p(\omega_{m_n}^{2d_1}).
\end{equation*}
Combining these bounds proves
\begin{equation}\label{eq_primitive_energy}
\sup_{d\in\mathcal D}\frac{m_n^{-1}\sum_{j=1}^{m_n}\omega_j^{2d}\|\mathcal{J}_{D,n}^R(\omega_j)\|^2}{S_n^0(d)}=O_p\{n(r_n^R)^2(m_n/n)^{2d_1}\}=o_p(1).
\end{equation}
Applying the same argument separately to the centered density-estimation and population-floor perturbations, and projecting onto the unit direction $v_1$, gives
\begin{equation*}
\mathcal E_{1,n}^R=O_p\{n(\delta_n/c_n)^2(m_n/n)^{2d_1}\},\qquad
\mathcal E_{2,n}^R=\begin{cases}
0, & R=R_1,\\
O_p\{n(\epsilon_n/c_n)^2(m_n/n)^{2d_1}\}, & R\in\{R_2,R_3\}.
\end{cases}
\end{equation*}
Condition~\eqref{eq_primitive_rate} therefore verifies Assumption~\ref{ass:generic}, and Theorem~\ref{thm:consistency} gives admissibility and consistency. The argument for the density-estimation component uses only $\delta_n/c_n\to0$, and hence also supplies the bound used in Remark~\ref{rem_separate_floor} when the floor contribution is verified separately.
\end{proof}

\begin{proof}[\textbf{Proof of Proposition~\ref{prop_direction}}]
Put $\Delta_n=\widehat v_1^R-\operatorname{sgn}(\langle\widehat v_1^R,v_1\rangle)v_1$, so that $\|\Delta_n\|=o_p(1)$.
After aligning the sign of the known-direction score, the additional Fourier error is $\langle\widehat{\mathcal{J}}_n^R(\omega_j),\Delta_n\rangle$.
The inequality $|\langle f,g\rangle|\leq\|f\|\|g\|$ gives
\begin{equation*}
    \sup_{d\in\mathcal D}
    \frac{m_n^{-1}\sum_{j=1}^{m_n}\omega_j^{2d}|\langle\widehat{\mathcal{J}}_n^R(\omega_j),\Delta_n\rangle|^2}{S_n^0(d)}
    \leq\|\Delta_n\|^2E_n^R=o_p(1).
\end{equation*}
Sign changes leave periodograms unchanged. The objective-comparison argument in Theorem~\ref{thm:consistency}, together with known-direction admissibility, gives objective equivalence. Consistency follows from the separation argument in Theorem~\ref{thm:consistency}.
\end{proof}

\begin{proof}[\textbf{Details on Remark~\ref{rem_lw_clt}}]
Put $a_n=\sqrt n\,r_n^R(m_n/n)^{d_1}$, so the rate condition in Remark~\ref{rem_lw_clt} gives $\sqrt{m_n}a_n\to0$. Let $v_n$ be either $v_1$ or the sign-aligned $\widehat v_1^R$, and write $\mathcal L_{m_n}^{0,v_n}$ and $\mathcal L_{m_n}^{R,v_n}$ for the latent and reconstructed regularized criteria using this same direction. For $j=1,\ldots,m_n$, define $J_j^0=(2\pi n)^{-1/2}\sum_{t=1}^n\langle\Y_t,v_n\rangle\exp(\mathrm it\omega_j)$ and $J_j^R=(2\pi n)^{-1/2}\sum_{t=1}^n\langle\widehat\Y_t^R,v_n\rangle\exp(\mathrm it\omega_j)$, and put $B_j=J_j^R-J_j^0$. The periodogram difference satisfies $|J_j^0+B_j|^2-|J_j^0|^2=|B_j|^2+2\operatorname{Re}\{J_j^0\overline{B_j}\}$.
The first and second derivatives of the two criteria can be expressed in terms of ratios of weighted periodogram sums; see \citet[Supplement, equations~(A.16), (A.19), and~(A.31)]{lrs2021_local_whittle}. We compare these sums using the preceding periodogram decomposition. By \eqref{eq_curve_error} and the argument for \eqref{eq_primitive_energy}, the terms involving $|B_j|^2$ contribute $O_p(a_n^2)$ to the derivative differences uniformly over a sufficiently small fixed neighborhood $U$ of $d_1$. The Cauchy--Schwarz inequality bounds the contribution of the cross terms by $O_p(a_n)$ on the same neighborhood. Consequently,
\begin{equation*}
\sup_{d\in U}\left|\partial_d^k\mathcal L_{m_n}^{R,v_n}(d)-\partial_d^k\mathcal L_{m_n}^{0,v_n}(d)\right|=O_p(a_n+a_n^2),\qquad k=1,2.
\end{equation*}
Under the maintained common scalar filter and regularity conditions, $\|v_n-v_1\|=o_p(1)$ allows the arguments establishing equations~(A.14)--(A.15) in the same supplement to apply to $\mathcal L_{m_n}^{0,v_n}$. Since $\sqrt{m_n}a_n\to0$, the preceding bounds preserve these derivative limits for $\mathcal L_{m_n}^{R,v_n}$. The Taylor expansion used in their proof of Theorem~1(ii) then yields \eqref{eq_lw_clt} and its known-direction counterpart.
\end{proof}

\section{Additional simulation studies}\label{sec:4}

\subsection{Sensitivity to the search interval under a unimodal baseline density}\label{sipp sect::unimodal additional simulation}

We use the same data-generating process as in Table~\ref{table::setting 1} and extend the search interval to $\mathcal D=[-0.49,0.49]$ as a sensitivity analysis. Table~\ref{table::setting 1 extended grid} reports nearly identical results to those in Table~\ref{table::setting 1}, indicating that the findings are robust to this wider search interval.

\begin{table}[!htb]
\centering
\renewcommand{\arraystretch}{1}
\caption{Aggregate Monte Carlo performance using extended search interval $\mathcal D=[-0.49,0.49]$. Each entry averages the cell-level absolute bias, RMSE, or MAE equally over all reported $n$ and $d_1$ cells; each cell contains 200 replications. Panel A uses the known $v_1$, whereas Panel B estimates $v_1$. Bias denotes average absolute bias.}\label{table::setting 1 extended grid}
\fontsize{7.3pt}{9.4pt}\selectfont
\setstretch{1.0}
\tabcolsep 0.25in
\begin{tabular}{@{}clrrrrrr@{}}
\toprule
& & \multicolumn{3}{c}{Panel A: Known $v_1$} & \multicolumn{3}{c}{Panel B: Estimated $v_1$} \\
\cmidrule(lr){3-5}\cmidrule(lr){6-8}
$N$     & Method & Bias & RMSE & MAE & Bias & RMSE & MAE \\
\midrule
100     & $R_0$ & .074 & .115 & .095 & .078 & .118 & .098 \\
        & $R_1$ & .047 & .096 & .077 & .044 & .094 & .075 \\
        & $R_2$ & .030 & .088 & .069 & .024 & .082 & .065 \\
        & $R_3$ & .033 & .089 & .070 & .027 & .083 & .066 \\
\addlinespace
200     & $R_0$ & .061 & .107 & .087 & .070 & .112 & .092 \\
        & $R_1$ & .033 & .090 & .071 & .030 & .087 & .069 \\
        & $R_2$ & .020 & .084 & .066 & .017 & .080 & .063 \\
        & $R_3$ & .024 & .085 & .067 & .020 & .080 & .064 \\
\addlinespace
400     & $R_0$ & .047 & .098 & .078 & .059 & .106 & .086 \\
        & $R_1$ & .022 & .085 & .067 & .018 & .083 & .065 \\
        & $R_2$ & .012 & .082 & .065 & .014 & .079 & .063 \\
        & $R_3$ & .016 & .083 & .065 & .016 & .079 & .063 \\
\bottomrule
\end{tabular}
\end{table}

\begin{figure}[!htb]
\centering
\includegraphics[width=0.9\linewidth]{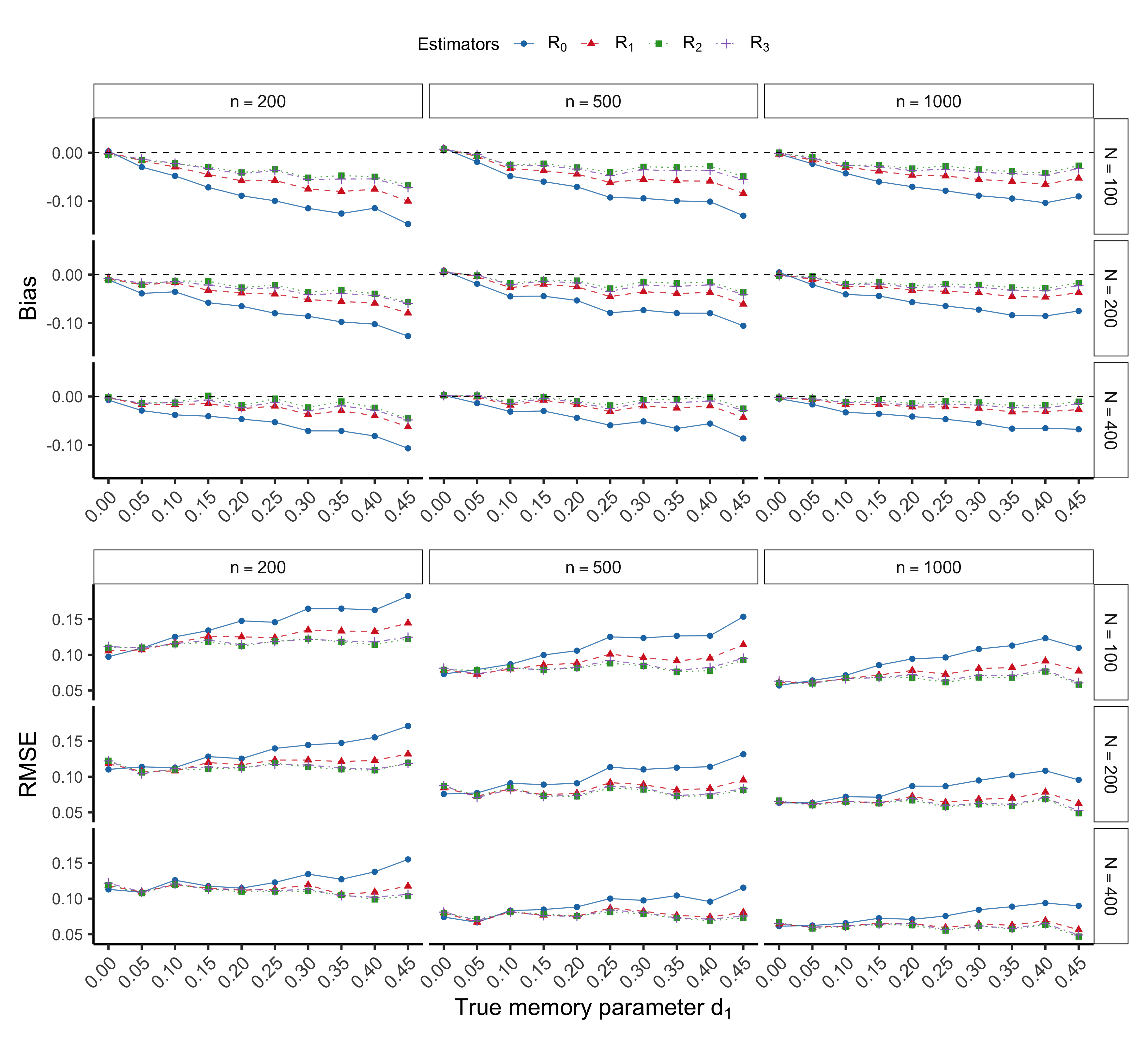}
\caption{\small Bias and RMSE of estimators $R_0$, $R_1$, $R_2$ and $R_3$, for the known $v_1$ setting with sample size $n\in\{200,500,1000\}$, the within-curve sample size $N\in \{100,200,400\}$, and the true long memory parameter $d_1\in\{0,0.05,\cdots,0.45\}$. The search interval is $\mathcal D=[-0.49,0.49]$.}\label{fig::sim est known extended grid}
\end{figure}

\begin{figure}[!htb]
\centering
\includegraphics[width=0.9\linewidth]{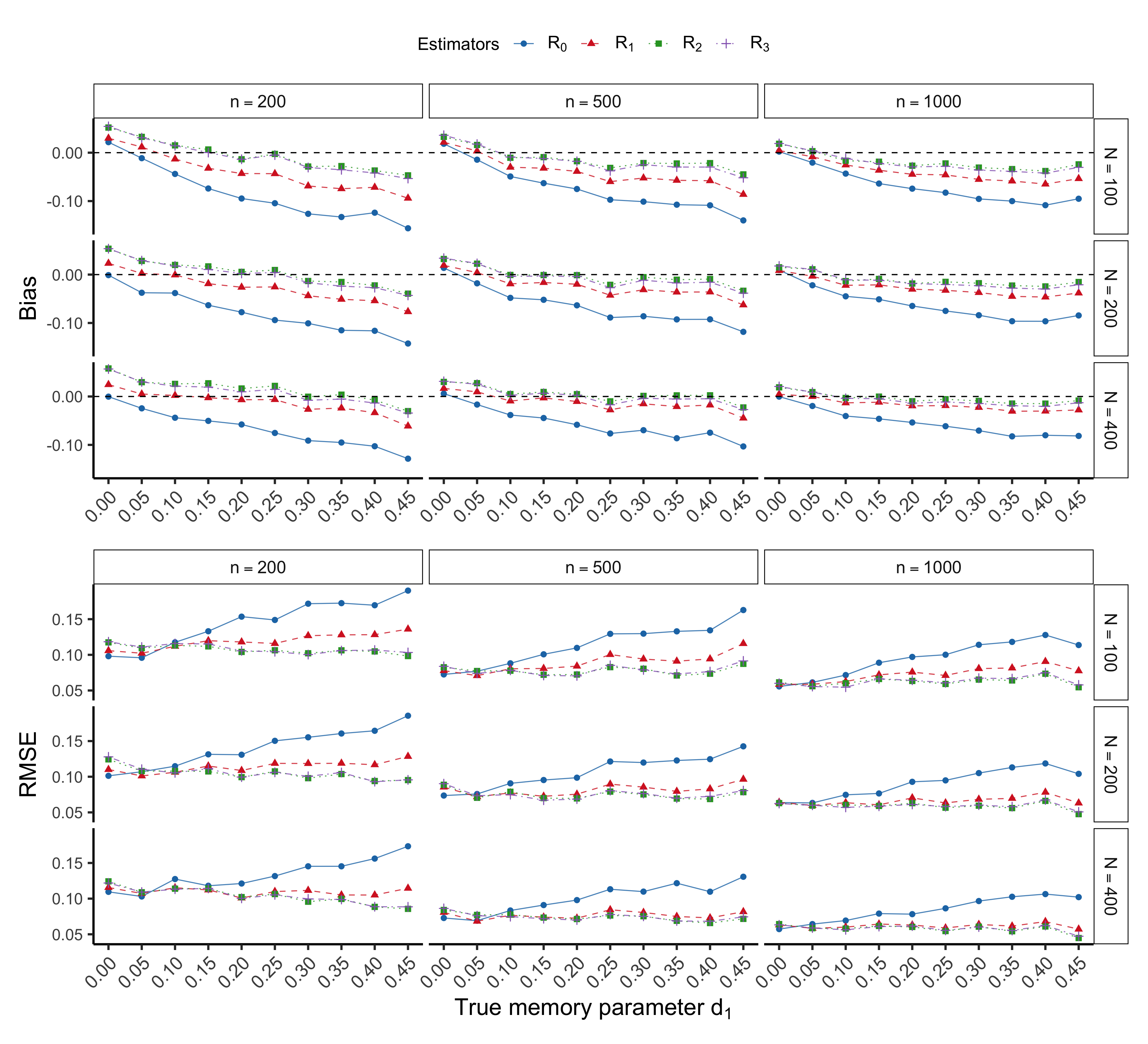}
\caption{\small Bias and RMSE of estimators $R_0$, $R_1$, $R_2$ and $R_3$, for the estimated $v_1$ setting with sample size $n\in\{200,500,1000\}$, the within-curve sample size $N\in \{100,200,400\}$, and the true long memory parameter $d_1\in\{0,0.05,\cdots,0.45\}$. The search interval is $\mathcal D=[-0.49,0.49]$.}\label{fig::sim est est extended grid}
\end{figure}

\begin{figure}[!htb]
\centering
\includegraphics[width=0.9\linewidth]{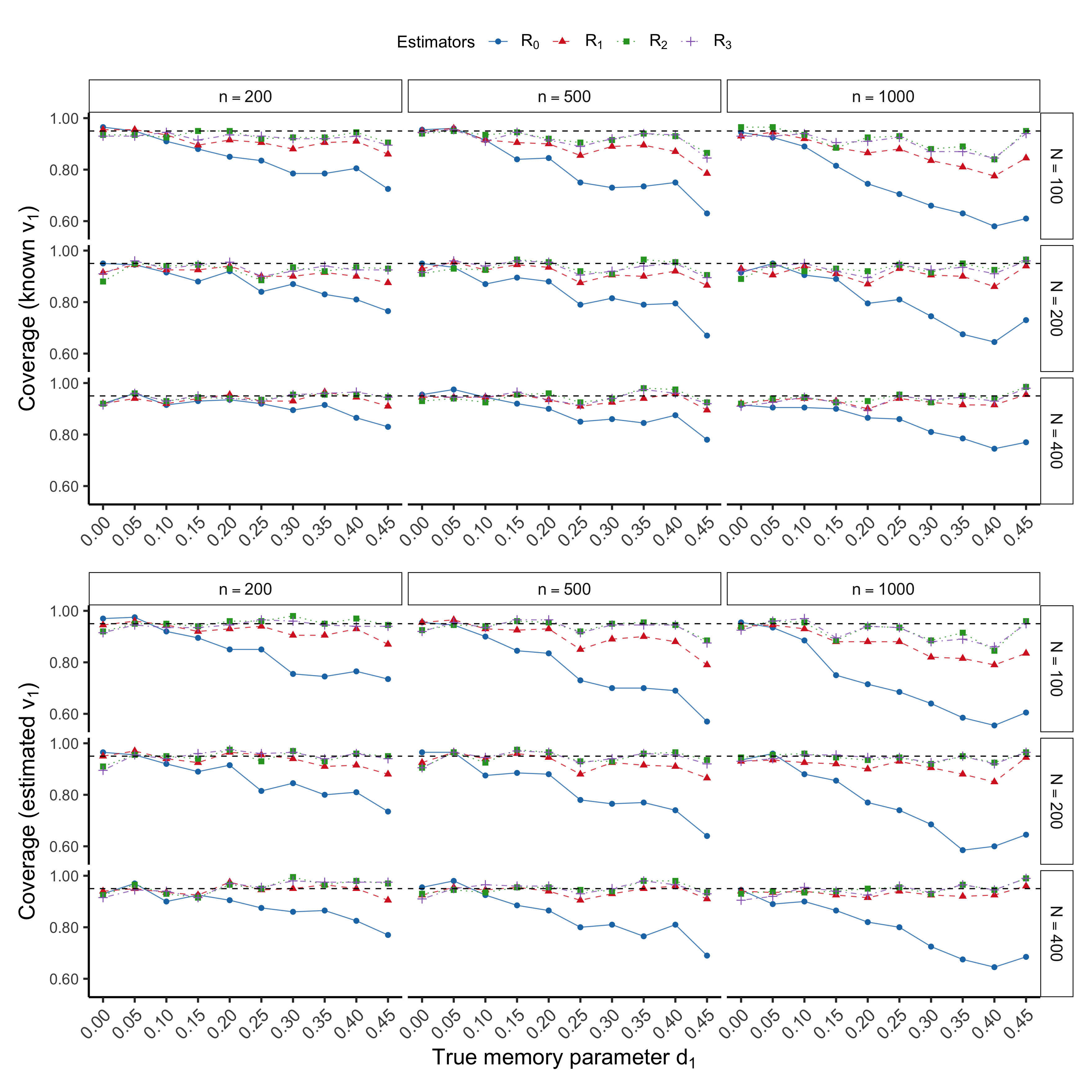}
\caption{\small Coverage produced by estimators $R_0$, $R_1$, $R_2$ and $R_3$, for either the known or estimated $v_1$ setting with sample size $n\in\{200,500,1000\}$, the within-curve sample size $N\in \{100,200,400\}$, and the true long memory parameter $d_1\in\{0,0.05,\cdots,0.45\}$. The horizontal dashed line stands for the nominal coverage probability of $0.95$. The search interval is $\mathcal D=[-0.49,0.49]$.}\label{fig::sim coverage known extended grid}
\end{figure}

Moreover, Figures~\ref{fig::sim est known extended grid}, \ref{fig::sim est est extended grid}, and \ref{fig::sim coverage known extended grid} also present identical patterns to those observed in Figures~\ref{fig::sim est known}, \ref{fig::sim est est}, and \ref{fig::sim coverage unimodel}, respectively.

\subsection{Uniform baseline density setting}

We consider an additional simulation setting to examine memory estimation when the baseline density does not have sparse tails. Compared to the data generation process in Section~\ref{sec:3.1}, the change is to replace the concentrated baseline density with a uniform baseline $f_0(u)=1$. Table~\ref{table::setting uniform} reports average absolute bias, RMSE, and MAE, when using the true leading direction $v_{1}$, or estimating $v_1$ by PCA from the regularized clr curves under the uniform baseline setting.  Additionally, following the analysis in Section~\ref{sipp sect::unimodal additional simulation}, we consider the extended search interval $\mathcal D=[-0.49,0.49]$ for the local Whittle estimator.

\begin{table}[!htb]
\centering
\renewcommand{\arraystretch}{1}
\caption{Uniform baseline setting: aggregate Monte Carlo performance. Each entry averages the cell-level absolute bias, RMSE, or MAE equally over all reported $n$ and $d_1$ cells; each cell contains 200 replications. Panel A uses the known $v_1$, whereas Panel B estimates $v_1$ by Bartlett long-run covariance PCA from each method's clr curves, with bandwidth $\lfloor n^{0.33}\rfloor$. Bias denotes average absolute bias.}
\label{table::setting uniform}
\fontsize{7.3pt}{9.4pt}\selectfont
\setstretch{1.0}
\setlength{\tabcolsep}{20pt}
\begin{tabular}{@{}clrrrrrr@{}}
\toprule
& & \multicolumn{3}{c}{Panel~A: Known $v_1$} & \multicolumn{3}{c}{Panel B: Estimated $v_1$} \\
\cmidrule(lr){3-5}\cmidrule(lr){6-8}
$N$ & Method & Bias & RMSE & MAE & Bias & RMSE & MAE \\
\midrule
100     & $R_0$ & .014 & .083 & .066 & .018 & .082 & .065 \\
        & $R_1$ & .012 & .082 & .065 & .016 & .080 & .064 \\
        & $R_2$ & .010 & .082 & .065 & .014 & .079 & .063 \\
        & $R_3$ & .011 & .082 & .065 & .015 & .080 & .064 \\
\addlinespace
200     & $R_0$ & .008 & .081 & .064 & .014 & .079 & .063 \\
        & $R_1$ & .008 & .081 & .064 & .015 & .078 & .063 \\
        & $R_2$ & .007 & .081 & .064 & .013 & .078 & .062 \\
        & $R_3$ & .008 & .081 & .064 & .015 & .078 & .063 \\
\addlinespace
400     & $R_0$ & .007 & .081 & .064 & .013 & .079 & .063 \\
        & $R_1$ & .006 & .080 & .064 & .017 & .078 & .063 \\
        & $R_2$ & .006 & .080 & .064 & .013 & .078 & .063 \\
        & $R_3$ & .006 & .080 & .064 & .017 & .078 & .063 \\
\bottomrule
\end{tabular}
\end{table}

From Table~\ref{table::setting uniform}, it is not surprising to find that the differences between the four procedures are much smaller under the uniform baseline setting than under the concentrated baseline setting in Section~\ref{sec:3.2}. Under the stationary score generation process, a uniform baseline density does not lead to the challenging density reconstruction scenario. Compared with Table~\ref{table::setting 1}, it suggests that the larger gains in Section~\ref{sec:3.2} are associated with the difficulty of reconstructing densities with sparsely sampled regions, as expected. Importantly, Table~\ref{table::setting uniform} indicates that, even under the uniform baseline setting, the proposed  $R_1$, $R_2$, and $R_3$ still have slightly better finite-sample performance.

\begin{figure}[!htb]
\centering
\includegraphics[width=0.9\linewidth]{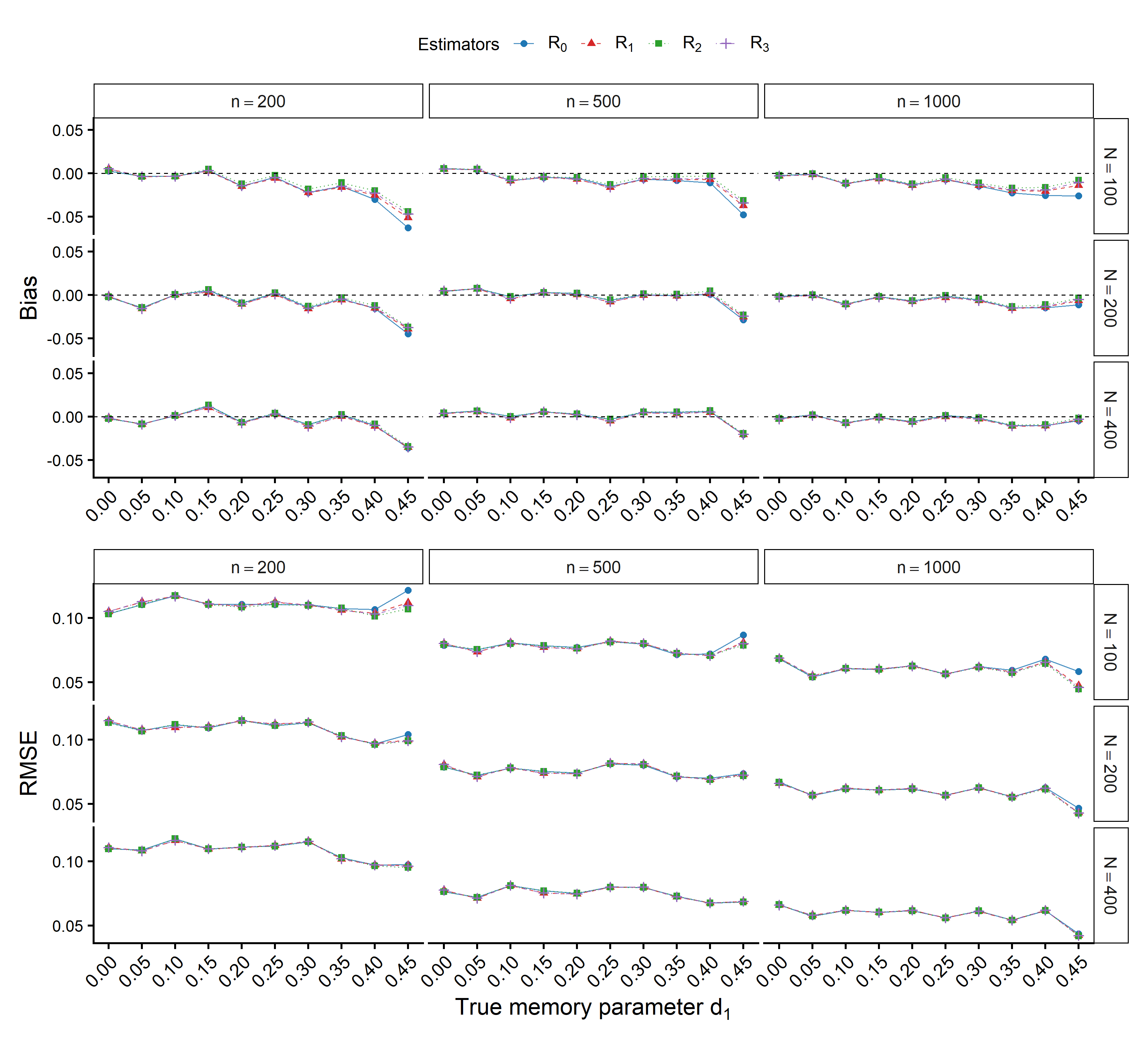}
\caption{\small Bias and RMSE of estimators $R_0$, $R_1$, $R_2$ and $R_3$, for the known $v_1$ setting with sample size $n\in\{200,500,1000\}$, the within-curve sample size $N\in \{100,200,400\}$, and the true long memory parameter $d_1\in\{0,0.05,\cdots,0.45\}$ under the uniform baseline setting.}\label{fig::sim est known uniform}
\end{figure}

\begin{figure}[!htb]
\centering
\includegraphics[width=0.9\linewidth]{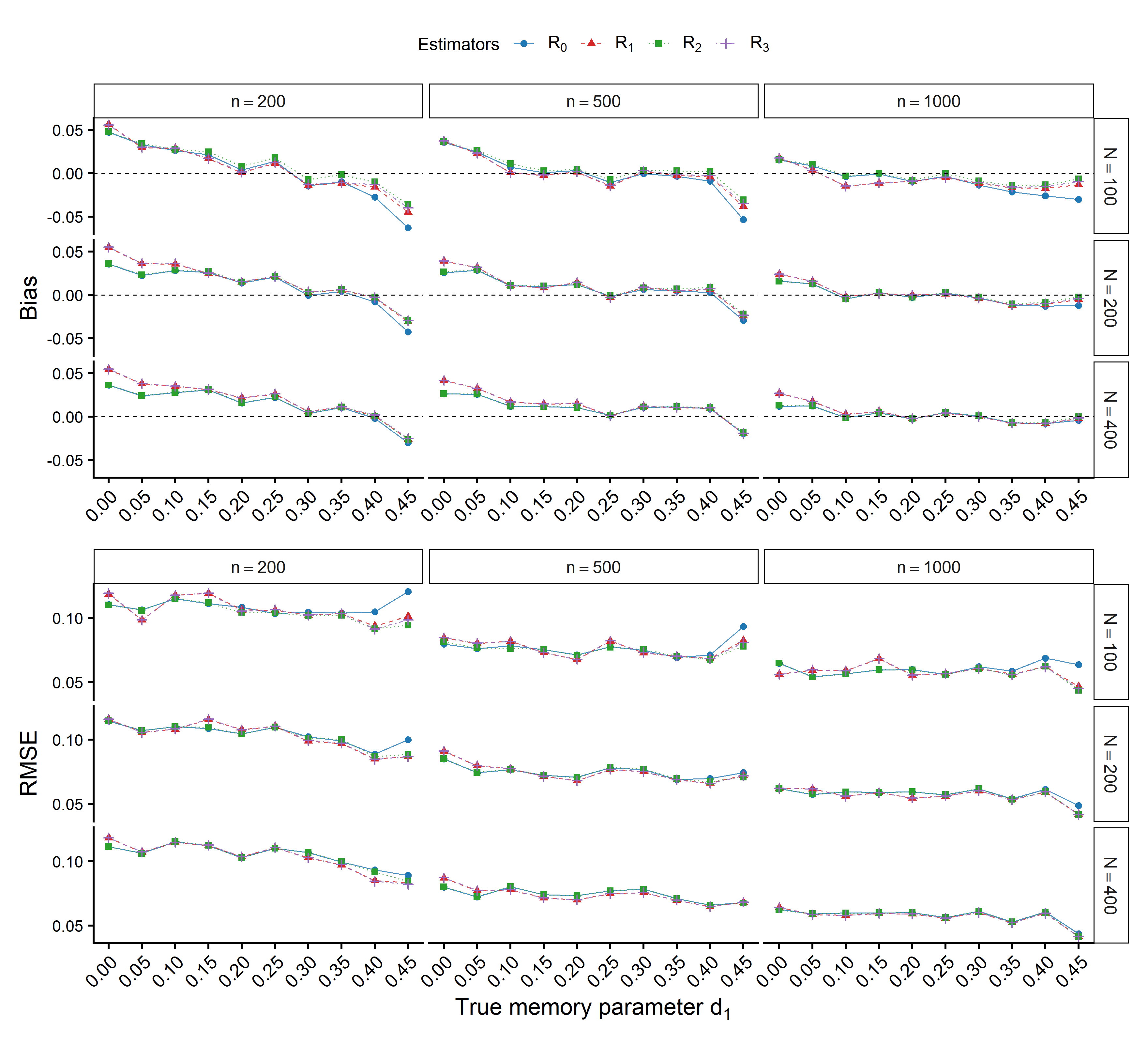}
\caption{\small Bias and RMSE of estimators $R_0$, $R_1$, $R_2$ and $R_3$, for the estimated $v_1$ setting with sample size $n\in\{200,500,1000\}$, the within-curve sample size $N\in \{100,200,400\}$, and the true long memory parameter $d_1\in\{0,0.05,\cdots,0.45\}$ under the uniform baseline setting.}\label{fig::sim est est uniform}
\end{figure}

Figures~\ref{fig::sim est known uniform} and~\ref{fig::sim est est uniform} report the bias and RMSE of estimators $R_0$, $R_1$, $R_2$ and $R_3$ for the known and estimated $v_1$ settings, respectively. They show that the bias and RMSE of the four procedures largely overlap, aligning with the findings in Table~\ref{table::setting uniform}. Figures~\ref{fig::sim est known uniform} and~\ref{fig::sim est est uniform} also show that, when $d_1=0.45$, $R_1$, $R_2$ and $R_3$ have smaller bias and RMSE compared to those of $R_0$, especially when the sample size $n$ or the within-curve sample size $N$ is small. Moreover, both the bias and RMSE of $R_0$, $R_1$, $R_2$ and $R_3$ become smaller when either the sample size $n$ increases or the within-curve sample size $N$ grows.

\begin{figure}[!htb]
\centering
\includegraphics[width=0.9\linewidth]{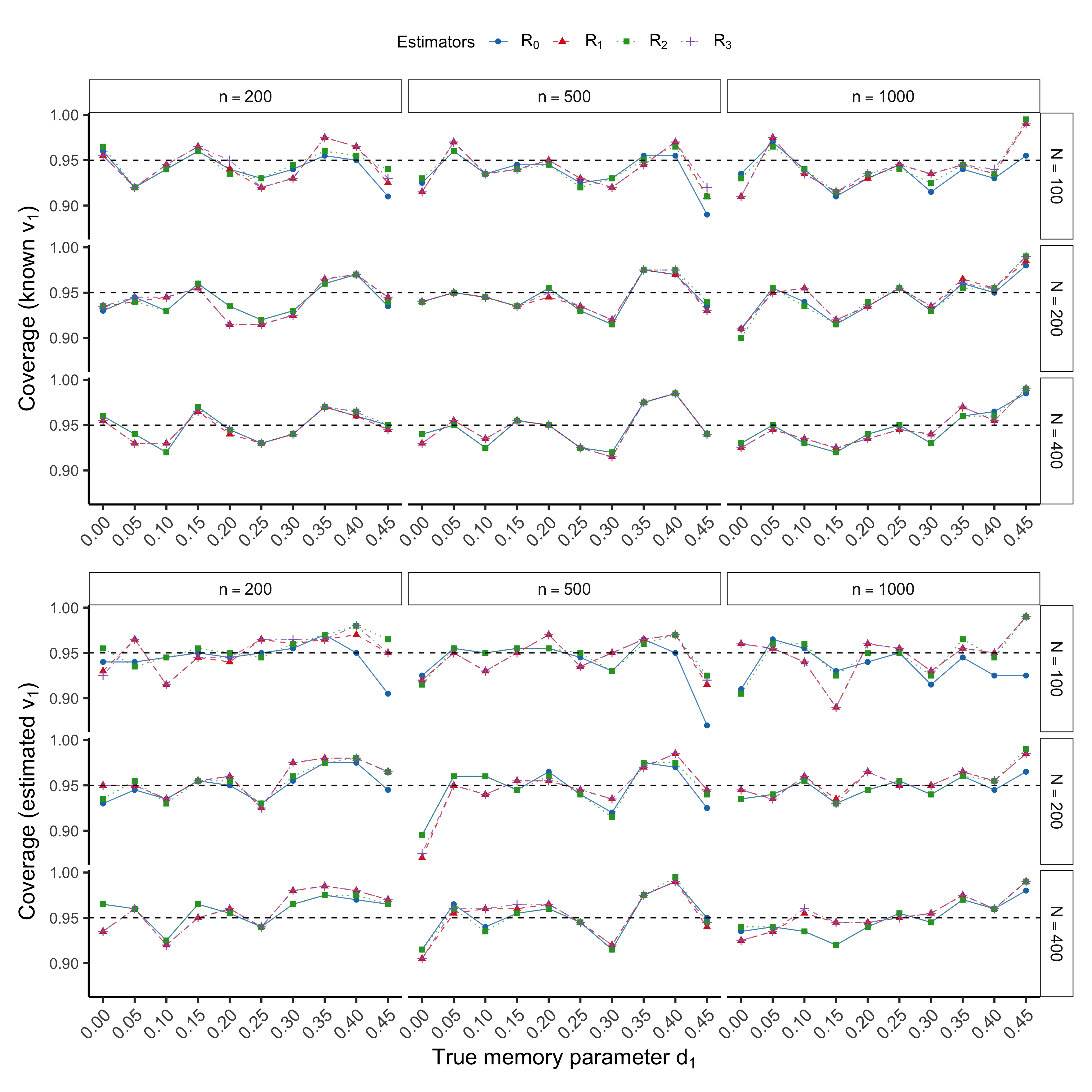}
\caption{\small Coverage produced by estimators $R_0$, $R_1$, $R_2$ and $R_3$, for either the known or estimated $v_1$ setting with sample size $n\in\{200,500,1000\}$, the within-curve sample size $N\in \{100,200,400\}$, and the true long memory parameter $d_1\in\{0,0.05,\cdots,0.45\}$ under the uniform baseline setting. The horizontal dashed line stands for the nominal coverage probability of $0.95$.}\label{fig::sim coverage known uniform}
\end{figure}

Figure~\ref{fig::sim coverage known uniform} provides a similar conclusion for the confidence interval. Different from Figure~\ref{fig::sim coverage unimodel}, the empirical coverage produced by $R_0$ has better finite-sample performance, indicating that the inferior performance observed in Figure~\ref{fig::sim coverage unimodel} is caused by the finite-sample bias.

\subsection{Nonstationary setting}

In this section, we evaluate the long memory parameter estimation under the nonstationary setting. We follow the same data generating process in Section~\ref{sec:3.1} with $d_1\in\{0,0.15,\cdots,0.9\}$. Here, $d_1>1/2$ corresponds to the nonstationary setting. We borrow the recursive integer-differencing procedure idea in Section~4 of \cite{lin2022not} to estimate $d_1$. Starting with the undifferenced scores, let $\widetilde d^{(q)}$ denote the local Whittle estimate obtained after $q$ integer differences, with $q=0$ corresponding to the original score series. If $\widetilde d^{(q)}<1-\gamma_n$ with $\gamma_n=0.4/\lfloor 1+n^{0.65}\rfloor$, the procedure terminates. Otherwise, we difference the current score series once more and recompute the local Whittle estimate. Writing $\widehat q$ for the first differencing order satisfying the stopping criterion, the final estimate is $\widehat d^{R,\mathrm{est}}=\widetilde d^{(\widehat q)}+\widehat q$. Note that when the initial estimate satisfies the stopping criterion, no differencing is performed.

Table~\ref{table::setting nonstationay} reports aggregate performance on $d_1\in\{0,0.15,\cdots,0.9\}$, which contains both stationary and nonstationary cases. The procedures $R_2$ and $R_3$, which involve the density-floor construction, yield the largest improvements. For example, with known $v_1$ and $N=100$, $R_2$ reduces average absolute bias of $R_0$ from $0.091$ to $0.025$, and RMSE from $0.150$ to $0.096$. Tail smoothing alone gives smaller reductions, with average absolute bias $0.078$ and RMSE $0.144$ for $R_1$. Combining the two regularizations yields performance close to flooring alone, with average absolute bias $0.031$ and RMSE $0.100$ for $R_3$.

\begin{table}[!htb]
\centering
\renewcommand{\arraystretch}{1}
\caption{Nonstationary setting: aggregate Monte Carlo performance. Each entry averages the cell-level absolute bias, RMSE, or MAE equally over all reported $n$ and $d_1$ cells; each cell contains 200 replications. Panel A uses the known $v_1$, whereas Panel B estimates $v_1$. Bias denotes average absolute bias.}\label{table::setting nonstationay}
\fontsize{7.3pt}{9.4pt}\selectfont
\setstretch{1.0}
\tabcolsep 0.25in
\begin{tabular}{@{}clrrrrrr@{}}
\toprule
& & \multicolumn{3}{c}{Panel~A: Known $v_1$} & \multicolumn{3}{c}{Panel~B: Estimated $v_1$} \\
\cmidrule(lr){3-5}\cmidrule(lr){6-8}
$N$ & Method & Bias & RMSE & MAE & Bias & RMSE & MAE \\
\midrule
100     & $R_0$ & .091 & .150 & .125 & .099 & .155 & .129 \\
        & $R_1$ & .078 & .144 & .115 & .083 & .146 & .117 \\
        & $R_2$ & .025 & .096 & .075 & .026 & .093 & .073 \\
        & $R_3$ & .031 & .100 & .078 & .031 & .097 & .076 \\
\addlinespace
200     & $R_0$ & .067 & .131 & .108 & .076 & .137 & .112 \\
        & $R_1$ & .054 & .123 & .097 & .061 & .125 & .099 \\
        & $R_2$ & .014 & .090 & .071 & .017 & .088 & .069 \\
        & $R_3$ & .017 & .091 & .071 & .021 & .088 & .070 \\
\addlinespace
400     & $R_0$ & .054 & .123 & .099 & .062 & .128 & .102 \\
        & $R_1$ & .040 & .113 & .088 & .049 & .116 & .090 \\
        & $R_2$ & .009 & .089 & .069 & .014 & .087 & .068 \\
        & $R_3$ & .010 & .088 & .069 & .016 & .087 & .068 \\
\bottomrule
\end{tabular}
\end{table}

\begin{figure}[!htb]
\centering
\includegraphics[width=0.9\linewidth]{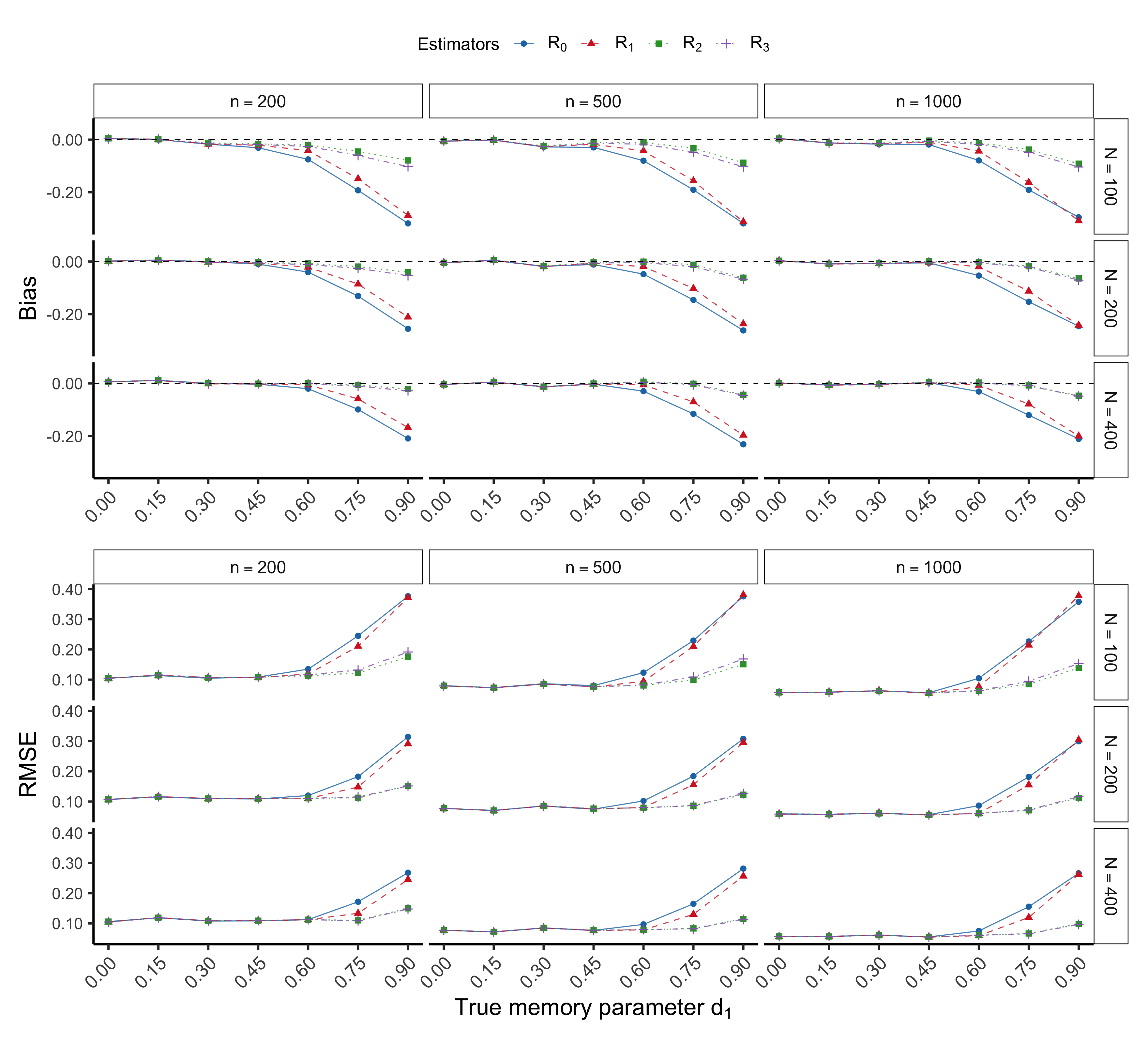}
\caption{\small Bias and RMSE of estimators $R_0$, $R_1$, $R_2$ and $R_3$, for the known $v_1$ setting with sample size $n\in\{200,500,1000\}$, the within-curve sample size $N\in \{100,200,400\}$, and the true long memory parameter $d_1\in\{0,0.15,\cdots,0.9\}$ under the nonstationay setting.}\label{fig::sim est known nonstationary}
\end{figure}

\begin{figure}[!htb]
\centering
\includegraphics[width=0.9\linewidth]{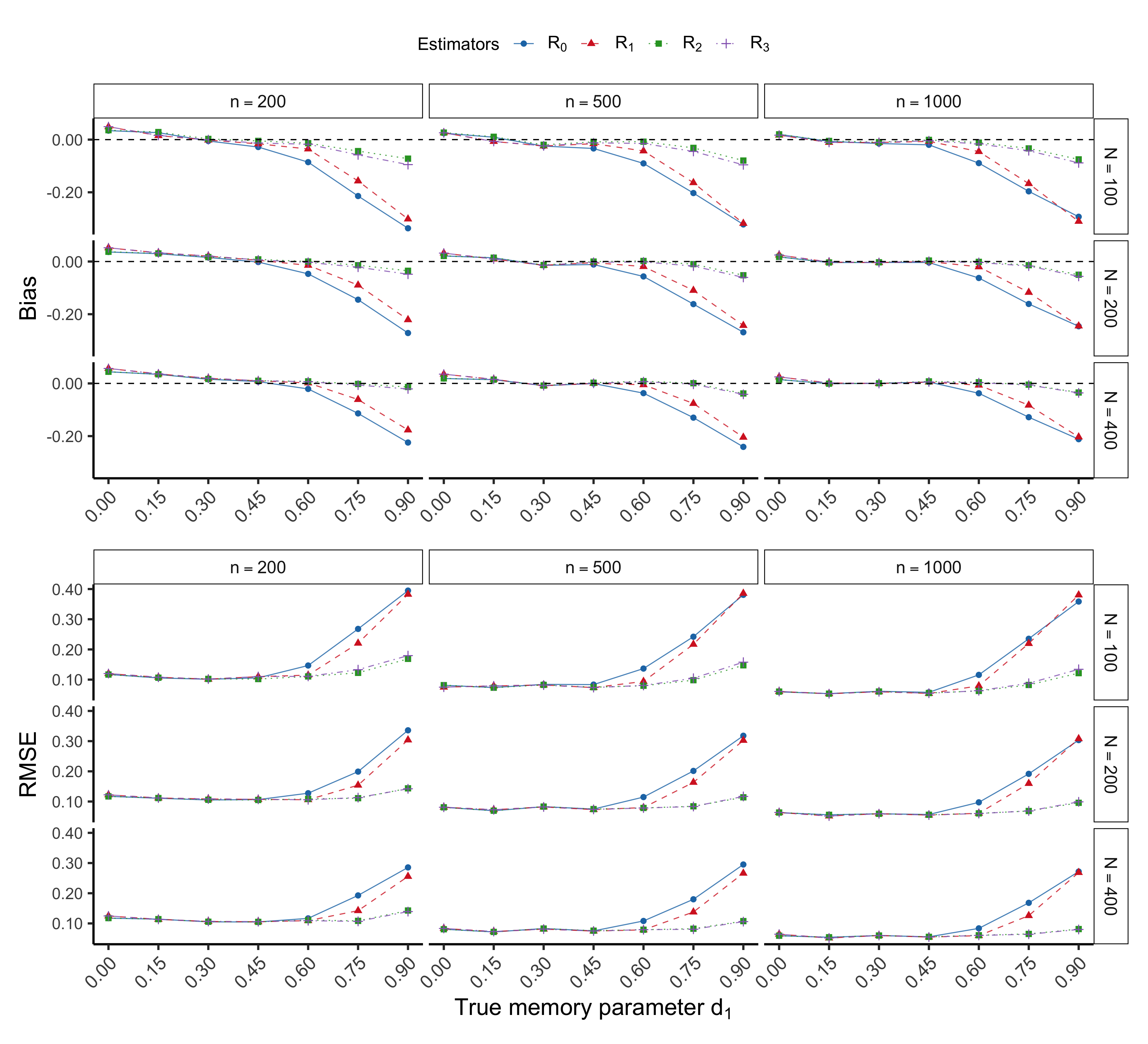}
\caption{\small Bias and RMSE of estimators $R_0$, $R_1$, $R_2$ and $R_3$, for the estimated $v_1$ setting with sample size $n\in\{200,500,1000\}$, the within-curve sample size $N\in \{100,200,400\}$, and the true long memory parameter $d_1\in\{0,0.15,\cdots,0.9\}$ under the nonstationay setting.}\label{fig::sim est est nonstationary}
\end{figure}

Figures~\ref{fig::sim est known nonstationary} and~\ref{fig::sim est est nonstationary} report the bias and RMSE of estimators $R_0$, $R_1$, $R_2$ and $R_3$ for the known and estimated $v_1$ settings, respectively. From Figures~\ref{fig::sim est known nonstationary} and~\ref{fig::sim est est nonstationary}, one can observe that, when $d_1\in\{0,0.15,\cdots,0.45\}$, corresponding to the stationary setting, all estimators $R_0$, $R_1$, $R_2$ and $R_3$ perform well, similar to the patterns in Figures~\ref{fig::sim est known uniform} and~\ref{fig::sim est est uniform}. This supports that the recursive integer-differencing procedure works well,

When $d_1\in\{0.45,0.6,\cdots,0.9\}$, the score processes are not stationary. For this scenario, $R_0$ and $R_1$ exhibit substantial downward bias and sharply increasing RMSE, while $R_2$ and $R_3$ are much less affected. This aligns with the expected difficulty of density reconstruction under the inverse-clr data-generating mechanism in the nonstationary setting. When we consider to use \eqref{eq:mc_clr_dgp}, for $d_1>1/2$, $\operatorname{Var}(\xi_{1t})\sim t^{2d_1-1}$ and this will lead to highly concentrated densities.

\clearpage
\bibliographystyle{agsm}
\bibliography{LRD_BHS}
\end{document}